\documentclass[letterpaper,twocolumn,10pt]{article}
\usepackage{usenix2019}
\usepackage{algorithm}
\usepackage{algorithmic}
\usepackage{graphicx}
\usepackage{amsmath}
\usepackage{subfigure}
\usepackage{caption}

\newcommand{\kvnet}{\textit{TurboKV}}
\newcommand{\kvfig}{Figure~}
\newcommand{\kvtable}{Table~}
\begin{document}

\date{}
\title{\Large \bf \kvnet: Scaling Up the Performance of Distributed Key-value Stores with In-Switch Coordination}
\author{
{\rm Hebatalla Eldakiky, David Hung-Chang Du, Eman Ramadan}\\
Department of Computer Science and Engineering\\
University of Minnesota - Twin Cities, USA\\
Emails: \{eldak002, du\}@umn.edu, eman@cs.umn.edu}
\maketitle

\begin{abstract}

The power and flexibility of software-defined networks lead to a programmable network infrastructure in which in-network computation can help accelerating the performance of applications. This can be achieved by offloading some computational tasks to the network. However, what kind of computational tasks should be delegated to the network to accelerate applications performance?
In this paper, we propose a way to exploit the usage of programmable switches to scale up the performance of distributed key-value stores. Moreover, as a proof-of-concept, we propose \kvnet, an efficient distributed key-value store architecture that utilizes programmable switches as: 1) partition management nodes to store the key-value store partitions and replicas information; and 2) monitoring stations to measure the load of storage nodes, this monitoring information is used to balance the load among storage nodes. We also propose a key-based routing protocol to route the search queries of clients based on the requested keys to targeted storage nodes. Our experimental results of an initial prototype show that our proposed architecture improves the throughput and reduces the latency of distributed key-value stores when compared to the existing architectures.
\end{abstract}

\section{Introduction}\label{sec:introduction}

Programmable switches in software-defined network promise flexibility and high throughput~\cite{pswitch1,pswitch2}. Recently, the Programming Protocol-Independent Packet Processor (P4)~\cite{p4} unleashes capabilities that give the freedom to create intelligent network nodes performing various functions. Thus, applications can boost their performance by offloading part of their computational tasks to these programmable switches to be executed in the network. Nowadays, programmable networks get a bigger foot in the data center doors. Google cloud started to use the programmable switches with P4Runtime~\cite{p4runtime} to build and control their smart networks~\cite{gcloud}. Some Internet Service Providers (ISPs), such as AT\&T, have already integrated programmable switches in their networks~\cite{att_info}. These switches can be controlled by network operators to adapt to the current network state and application requirements.
 
Recently, there has been an uptake in leveraging programmable switches to improve distributed systems, e.g., NetPaxos~\cite{netpaxos1,netpaxos2}, NetCache~\cite{netcache}, NetChain~\cite{netchain}, DistCache~\cite{distcache} and iSwitch~\cite{iswitch}. This uptake is due to the massive evolution on the capailities of these network switches, e.g., Tofino ASIC from Barefoot~\cite{tofino} which provides sub-microsecond per-packet processing delay with bandwidth up to 6.5 Tbs and throughput of few billions of packets processed per second and Tofino2 with bandwidth up to 12.8 Tbs. These systems use the switches to provide orders of magnitude higher throughput than the traditional server-based solutions. 

On the other hand, through the massive use of mobile devices, data clouds, and the rise of Internet of Things~\cite{internet_of_things}, enormous amount of data has been generated and analyzed for the benefit of society at a large scale. This data can be text, image, audio, video, etc., and is generated by different sources with un-unified structures. Hence, this data is often maintained in key-value storage, which is widely used due to its efficiency in handling data in key-value format, and flexibility to scale out without significant database redesign. Examples of popular key-value stores include Amazon's Dynamo~\cite{dynamo}, Redis~\cite{redisref}, RAMCloud~\cite{ramcloud}, LevelDB~\cite{leveldb} and RocksDB~\cite{rocksdb}. 

Such huge amount of data can not be stored in a single storage server. Thus, this data has to be partitioned across different storage instances inside the data center. The data paritions and their storage node mapping (directory information) are either stored on a single coordinator node, e.g., the master coordinator in distributed Google file system~\cite{gfs}, or replicated over all storage instances~\cite{cassendra,dynamo, redisref}. In the first approach, the master coordinator represents a single point of failure, and introduces a bottleneck in the path between clients and storage nodes; as all queries are directed to it to know the data location. Moreover, the query response time increases, and hence, the storage system performance decreases. In the second approach, where the directory information is replicated on all storage instances, there are two strategies that a client can use to select a node where the request will be sent to: \textit{server-driven coordination} and \textit{client-driven coordination}.

In server-driven coordination, the client routes its request through a generic load balancer that will select a node based on load information. The selected node acts like the coordinator for the client request. It answers the query if it has the data partition or forwards the query to the right instance where the data partition resides. In this strategy, the client neither has knowledge about the storage nodes nor needs to link any code specific to the key-value storage it contacts. Unfortunately, this strategy has a higher latency because it introduces additional forwarding step when the request coordinator is different from the node holding the target data.

In client-driven coordination, the client uses a partition-aware client library that routes requests directly to the appropriate storage node that holds the data. This approach achieves a lower latency compared to the server-driven coordination as shown in~\cite{dynamo}. In~\cite{dynamo}, the client-driven coordination approach reduces the latencies by more than 50\% for both $99.9^{th}$ percentile and average cases. This latency improvement is because the client-driven coordination eliminates the overhead of the load balancer and skips a potential forwarding step introduced in the server-driven coordination when a request is assigned to a random node. However, it introduces additional load on the client to periodically pickup a random node from the key-value store cluster to download the updated directory information to perform the coordination locally on its side. It also requires the client to equip its application with some code specific to the key-value store used.

Another challenge in maintaining distributed key-value stores is how to handle dynamic workloads and cope with changes in data popularity~\cite{workloadAnalysis,ycsb}. Hot data receives more queries than cold data, which leads to load imbalance between the storage nodes; some nodes are heavily congested while others become under-utilized. This results in a performance degradation of the whole system and a high tail latency. 

In this paper, we propose \kvnet~: a novel Distributed Key-Value Store Architecture that leverages the power and flexibility of the new generation of programmable switches. \kvnet~scales up the performance of the distributed key-value storage by offloading the partitions management and query routing to be carried out in network switches. \kvnet~uses a \textit{switch-driven coordination} which utilizes the programmable switches as: 1) partition management nodes to store and manage the directory information of key-value store; and 2) monitoring stations to measure the load of storage nodes, where this monitoring information is used to balance the load among storage nodes. 

\kvnet~adapts a hierarchical indexing scheme to distribute the directory information records inside the data center network switches. It uses a key-based routing protocol to map the requested key in the query packet from the client to its target storage node by injecting some information about the requested data in packet headers. The programmable switches use this information to decide where to send the packet to reach the target storage node directly based on the directory information records stored in the switches' data plane. This in-switch coordination approach removes the load of routing the requests from the client side in the client-driven coordination without introducing an additional forwarding step introduced by the coordination node in the server-driven coordination.

To achieve both reliability and high availability, \kvnet~replicates key-value pair partitions on different storage nodes. For each data partition, \kvnet~maintains a list of nodes that are responsible for storing the data of this partition. \kvnet~uses the chain replication model to guarantee strong data consistency between all partition replicas. In case of having a failing node, requests will be served with other available nodes in the partition replica list.

\kvnet~also handles load balancing by adapting a dynamic allocation scheme that utilizes the architecture of software-defined network~\cite{sdn_history,open_flow}. In our architecture, a logically centralized controller, which has a global view of the whole system~\cite{nox}, makes decisions to migrate/replicate some of the popular data items to other under-utilized storage nodes using monitoring reports from the programmable switches. Then, it updates the switches' data plane with the new indexing records. To the best of our knowledge, this is the first work to use the programmable switches as the request coordinator to manage the distributed key-value store's partition information along with the key-based routing protocol. Overall, our contributions in this paper are four-fold:

\begin{itemize}

  \item We propose the in-switch coordination paradigm, and design an indexing scheme to manage the directory information records inside the programmable switch along with protocols and algorithms to ensure the strong consistency of data among replica, and achieve the reliability and availability in case of having nodes failure. 

  \item We introduce a data migration mechanism to provide load balancing between the storage nodes based on the query statistics collected from the network switches.

  \item We propose a hierarchical indexing scheme based on our proposed rack scale switch coordinator design to scale up \kvnet~to multiple racks inside the existing data center network architecture.   

  \item We implemented a prototype of \kvnet~using P4 on top of the simple software switch architecture BMV2~\cite{bmv2}. Our experimental results show that our proposed architecture improves the throughput and reduces the latency for the distributed key-value stores.
\end{itemize}

The remaining sections of this paper are organized as follows. Background and motivation is discussed in Section~\ref{sec:background}. Section~\ref{sec:architecture} provides an overview of the \kvnet~architecture, while the detailed design of \kvnet~is presented in Section~\ref{sec:data_design} and Section~\ref{sec:control_design}. Section~\ref{sec:scaling_up} discusses how to scale up \kvnet~inside the data center networks. \kvnet~implementation is discussed in Section~\ref{sec:implementation}, while Section~\ref{sec:results} gives an experimental evidence and analysis of \kvnet. Section~\ref{relatedwork} provides a short survey about the related work to us, and finally, the paper is concluded in Section~\ref{sec:conclusion}.

\section{Background and Motivation}\label{sec:background}

\subsection{Preliminaries on Programmable Switches}

Software-Defined Network (SDN) simplifies network devices by introducing a logically centralized controller (control plane) to manage simple programmable switches (data plane). SDN controllers set up forwarding rules at the programmable switches and collect their statistics using OpenFlow APIs~\cite{open_flow}. As a result, SDN enables efficient and fine-grained network management and monitoring in addition to allowing independent evolution of the controller and programmable switches.  

Recently, P4~\cite{p4} has been introduced to enrich the capabilities of network devices by allowing developers to define their own packet formats and build the processing graphs of these customized packets. P4 is a programming language designed to program parsing and processing of user-defined packets using a set of match/action tables. It is a target-independent language, thus a P4 compiler is required to translate P4 programs into target-dependent switch configurations. 

\begin{figure}
 \centering 
 \includegraphics[width=\linewidth]{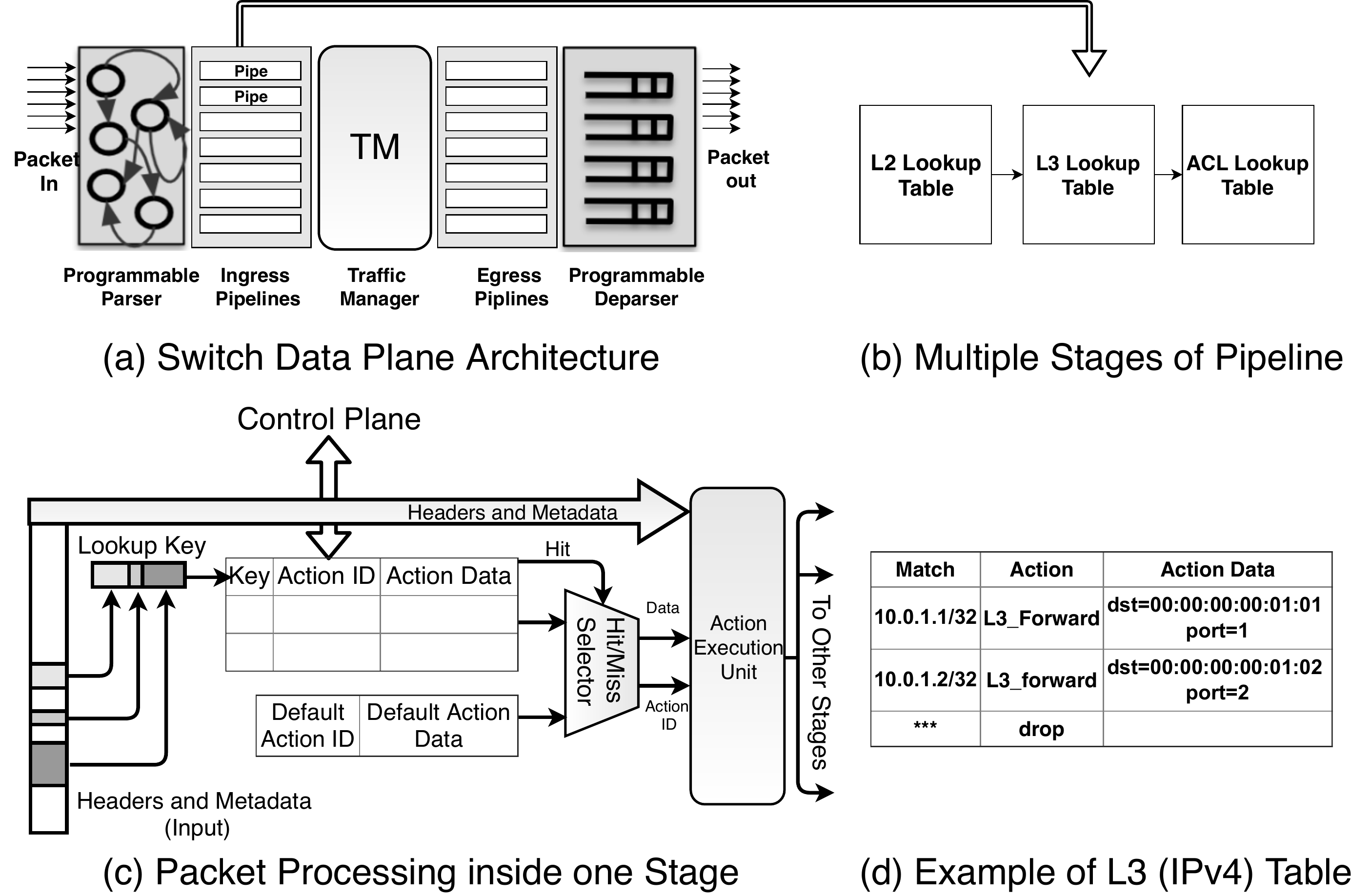}
 \caption{Primelinries on Programmable Switches}
 \label{fig:sw_arch}
\end{figure}

\kvfig\ref{fig:sw_arch}(a) represents the five main data plane components for most of modern switch ASICs. These components include programmable parser, ingress pipeline, traffic manager, egress pipeline and programmable deparser. When a packet is first received by one of the ingress ports, it goes through the programmable parser. The programmable parser, as shown in \kvfig\ref{fig:sw_arch}(a), is modeled as a simple deterministic state machine, that consumes packet data and identifies headers that will be recognized by the data plane program. It makes transitions between states typically by looking at specific fields in the previously identified headers. For example, after parsing the Ethernet header in the packet, the next state will be determined based on the Ethertype field, whether it will be reading a VLAN tag, an IPv4 or IPv6 header.

After the packet is processed by the parser and decomposed into different headers, it goes through one of the ingress pipes to enter a set of match-action processing stages as shown in \kvfig\ref{fig:sw_arch}(b). In each stage, a key is formed from the extracted data and matched against a table that contains some entries. These entries can be added and removed through the control plane. If there is a match with one of the entries, the corresponding action will be executed, otherwise a default action will be executed. After the action execution, the packet with all the updated headers from this stage enters the next stages in the pipeline. \kvfig\ref{fig:sw_arch}(c) illustrates the processing inside a pipeline stage, while \kvfig\ref{fig:sw_arch}(d)  
shows an example IPv4 table that picks an egress port based on destination IP address. The last rule drops all packets that do not match any of the IP prefixes, where this rules corresponds to the default action.

After the packet finishes all the stages in the ingress pipeline, it is queued and switched by the traffic manager for an egress pipe for further processing. At the end, there is a programmable deparser that performs the reverse operation of the parser. It reassembles the packet back with the updated headers so that the packet can go back onto the wire to the next hop in the path to its destination. Developers of P4 programs should take into consideration some constraints when designing a P4 Program. These constrains include: (i) the number of pipes, (ii) the number of stages and ports in each pipe, and (iii) the amount of memory used for prefix matching and data storage in each stage.

\subsection{Why In-Switch Coordination?}

We utilize the programmable switches to scale up the performance of distributed key-value stores with the in-switch coordination because of three reasons. First, programmable switches have become the backbone technology used in modern data centers or rack-scale clusters that allow developers to define their own functions for packet processing and provide flexibility to program the hardware. For example, Google uses the programmable switches to build and control their data centers networks~\cite{gcloud}.       

Second, request latency is one of the most important factors that affect the performance of all key-value stores. This latency is introduced through the multiple hops that the request traverses before arriving to its final destination, in addition to the processing time to fetch the desired key-value pair(s) from that destination. When the key-value store is distributed among several storage nodes, the request latency increases because of the latency introduced to figure out the location of the desired key-value pair(s) before fetching them from that location. This process is referred to as \textit{partition management and request coordination}, which can be organized by the storage nodes (server-based coordination) or by the client (client-based coordination). 

\begin{figure}
\centering 
\subfigure[In-Switch (2 hops)]{\includegraphics[width=0.45\linewidth]{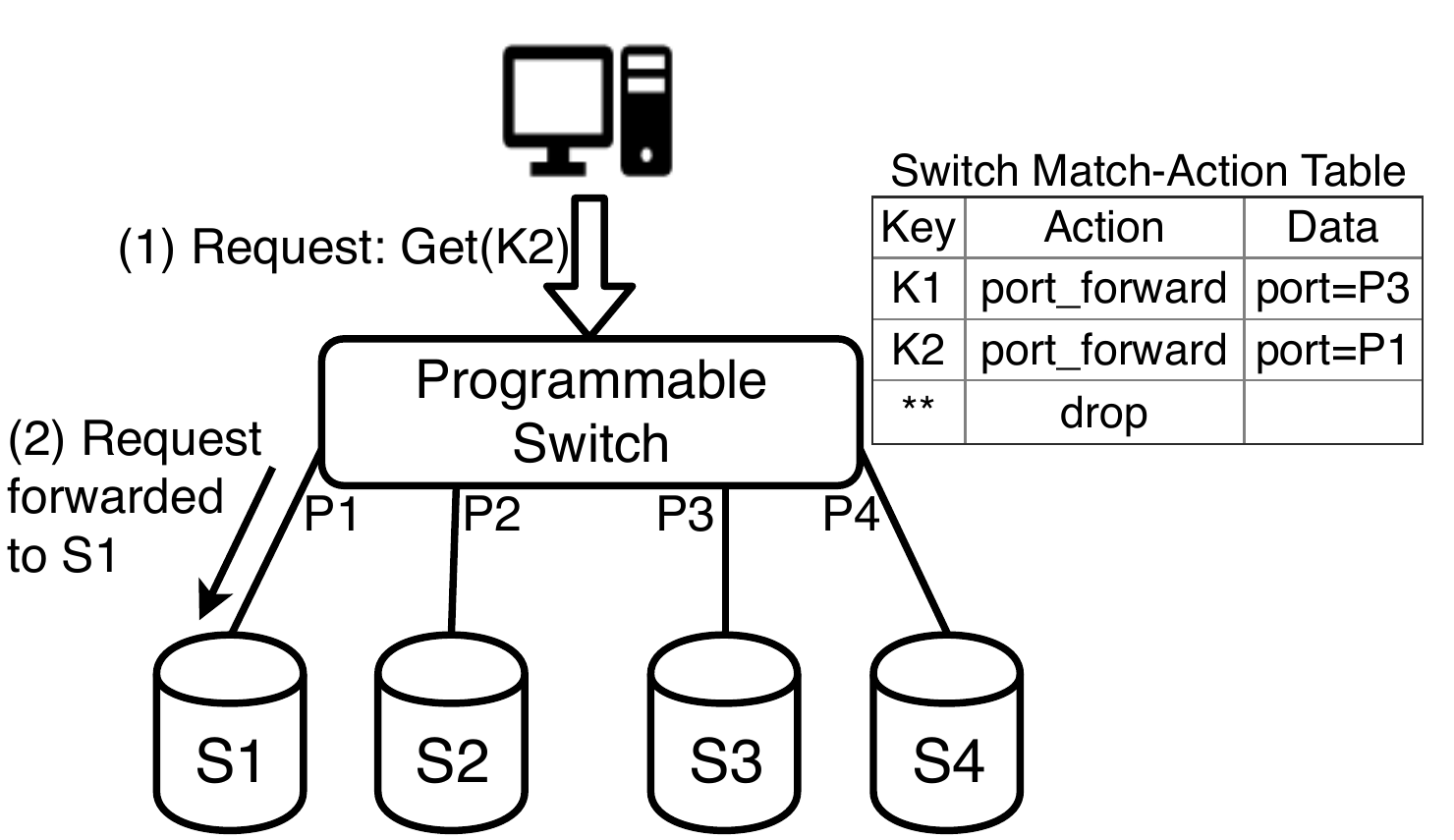}}
\subfigure[Server-based (4 hops)]{\includegraphics[width=0.45\linewidth]{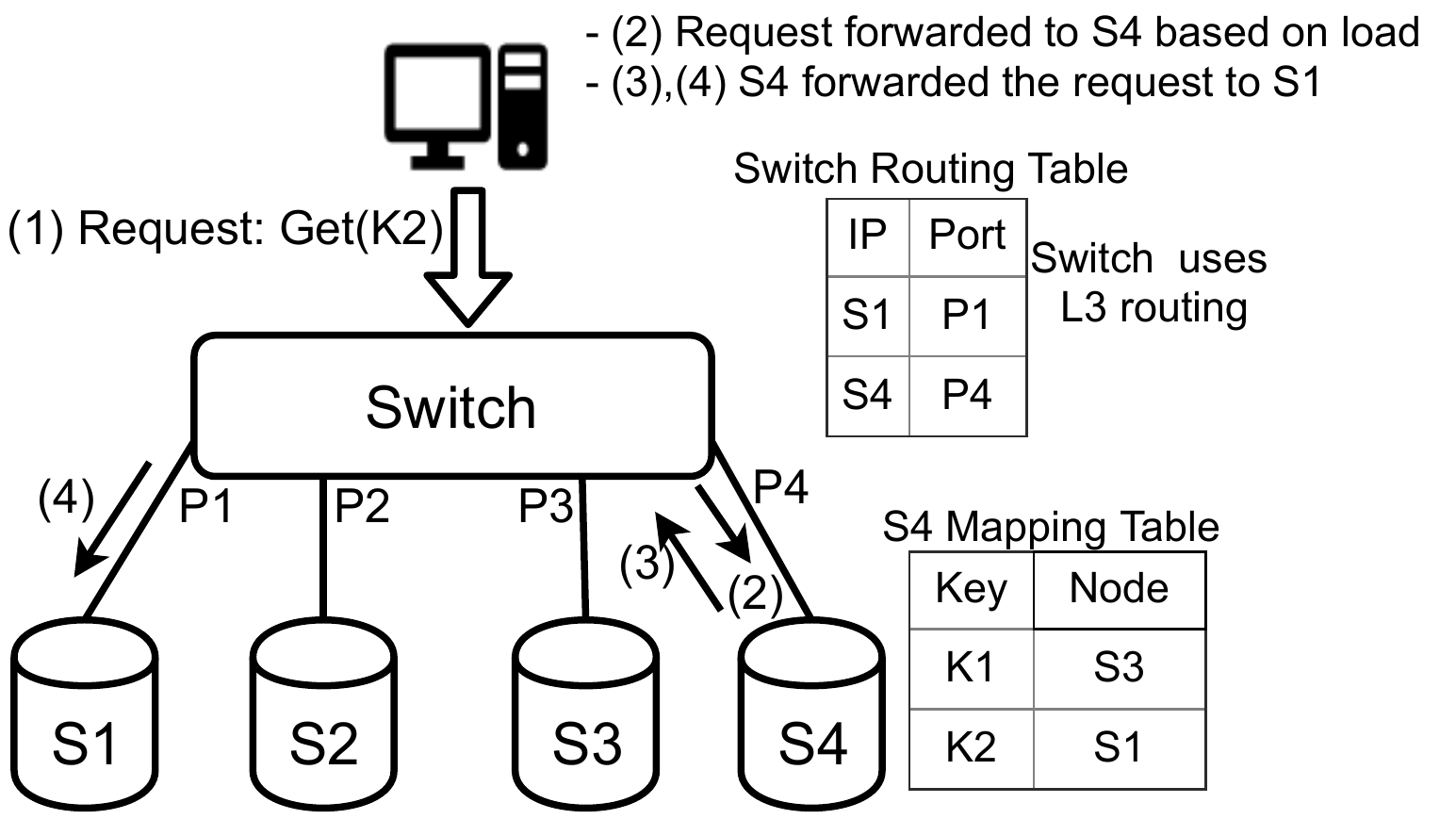}}
\caption{Request Coordination Models}
\vspace{-15 px}
\label{fig:serv_sw}
\end{figure}

Because client requests already pass through network switches to arrive at their target, offloading the partition management and query routing to be carried out in the network switches with the in-switch coordination approach will reduce the latency introduced by the server-based coordination approach; the number of hops that the request will travel from the client to the target storage node will be reduced as shown in \kvfig\ref{fig:serv_sw}. In-switch coordination also removes the load from the client in the client-based coordination approach by making the programmable switch manage all the information for the request routing.

Third, the partition mangement and request coordination are communication-bounded rather than being computation-bounded. So, the massive evolution in the capailities of the network switches, which provide orders of magnitude higher throughput than the highly optimized servers, makes them the best option to be used as partition mangement and request cordination nodes. For example, Tofino ASIC from Barefoot~\cite{tofino} provides few billions of packets processed per second with 6.5 Tbps bandwidth. Such performance is orders of magnitude higher than NetBricks~\cite{netbricks} which processes millions of packets per second and has 10-100 Gbps bandwidth~\cite{netchain}.  

\section{\kvnet~Architecture Overview}\label{sec:architecture}

\begin{figure}[t]
 \centering 
 \includegraphics[width=0.8\linewidth]{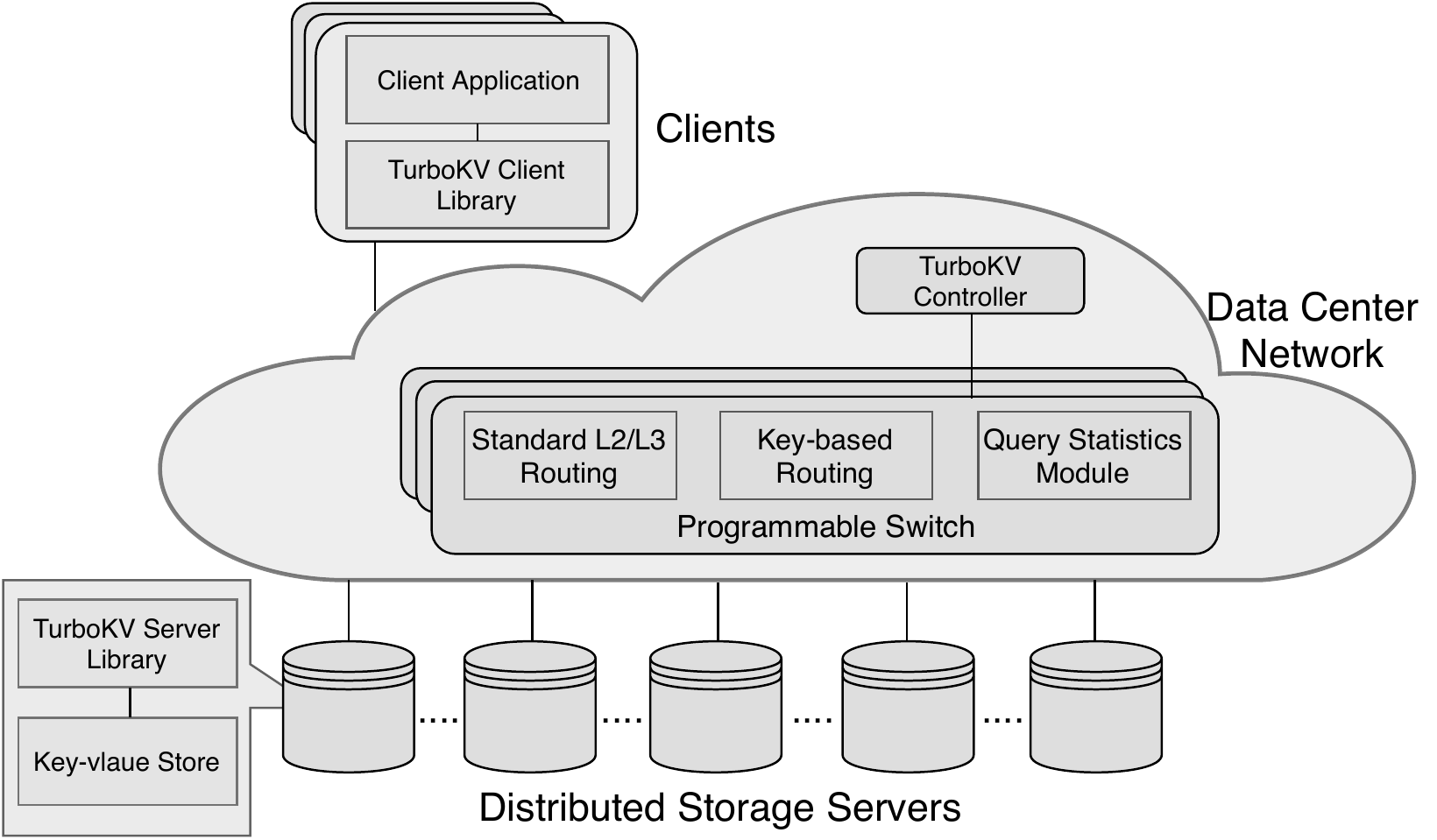}
 \caption{\kvnet~Architecture}
 \vspace{-10 px}
 \label{fig:KVNet_architecture}
\end{figure}

\kvnet~is a new architecture of the future distributed key-value stores that leverages the capability of programmable switches. \kvnet~uses an in-switch coordination approach to maintain the partition management information (directory information) of distributed key-value stores, and route clients' search queries based on their requested keys to the target storage nodes. \kvfig~\ref{fig:KVNet_architecture} shows the architecture of \kvnet~which consists of the programmable switches, controller, storage nodes, and system clients. 

\vspace{10 px}
\noindent\textbf{Programmable Switches.}
Programmable switches are the essential component in our new proposed architecture. We augment the programmable switches with a key-based routing approach to deliver \kvnet~query packets to the target key-value storage node. We leverage match-action tables and switch's registers to design the in-switch coordination where the partition management information will be stored on the path from the client to the storage node. This directory information represents the routing information to reach one of the storage nodes (Section~\ref{sec:data_plane_design}). Following this approach, the programmable switches act as request coordinator nodes that manage the data partitions and route requests to target storage nodes. 

In addition to using the key-based routing module, other packets are processed and routed using the standard L2/L3 protocols which makes \kvnet~compatible with other network functions and protocols (Section~\ref{sec:network_design}). Each programmable switch has a query statistics module to collect information about each partition's popularity to estimate the load of storage nodes (Section~\ref{sec:query_statistics}). This is vital to make data migration decisions to balance the load among storage nodes especially to handle dynamic workloads where the key-value pair popularity changes overtime.  

\vspace{10 px} 
\noindent\textbf{Controller.}
The controller is primarily responsible for system reconfigurations including (a)~achieving load balancing between the distributed storage nodes (Section~\ref{sec:query_statistics}), (b)~handling failures in the storage network (Section~\ref{sec:failure_handling}), and (c)~ updating each switch's match-action tables with the new location of data. The controller receives periodic reports form switches about the popularity of each data partition. Based on these reports, it decides to migrate/replicate part of the popular data to another storage node to achieve load balancing. Through the control plane, the controller updates the match-action tables in the switches with the new data locations. \kvnet~controller is an application controller that is different from the network controller in SDN, and it does not interfere with other network protocols or functions managed by the SDN controller. Our controller only manages the key-range based routing and data migrations and failures associated with them. Both controllers can be co-located on the same server, or on different servers.

\vspace{10 px} 
\noindent\textbf{Storage Nodes.} They represent the location where the key-value pairs reside in the system. The key-value pairs are partitioned among these storage nodes (Section~\ref{sec:data_partitioning}). Each storage node runs a simple shim that is responsible for reforming \kvnet~query packets to API calls for the key-value store, and handling \kvnet~controller's data migration requests between the storage nodes. This layer makes it easy to integrate our design with existing key-value stores without any modifications to the storage layer.       

\vspace{10 px}
\noindent\textbf{System Clients.} 
\kvnet~provides a client library which can be integrated with the client applications to send \kvnet~packets through the network, and access the key-value store without any modifications to the application. Like other key-value stores such as LevelDB\cite{leveldb} and RocksDB\cite{rocksdb}, the library provides an interface for all key-value pair operations that is responsible for constructing the \kvnet~packets and translates the reply back to the application.

\section{\kvnet~Data Plane Design}\label{sec:data_design}

The data plane provides in-switch coordination model for the key-value stores. In this model, all partition management information and query routing are managed by the switches. \kvfig\ref{fig:switch_pipeline} represents the whole pipeline that the packet traverses inside the switch before being forwarded to the right storage node. In this section, we discuss how the switch data plane supports these functions.

\subsection{On-Switch Partition Management}\label{sec:data_plane_design}

\subsubsection{Data Partitioning}\label{sec:data_partitioning}

In large-scale key-value stores, data is partitioned among storage nodes. \kvnet~supports two different partitioning techniques. Applications can choose any of these two partitioning techniques according to their needs.

\vspace{10 px}
\noindent\textbf{Range Partitioning.} In range partitioning, the whole key space is partitioned into small disjoint sub-ranges.
Each sub-range is assigned to a node (or multiple nodes depending on the replication factor). The data with keys fall into a certain sub-range is stored on the same storage node that is responsible for this sub-range. The key itself is used to map the user request to one of the storage nodes. Each storage node can be responsible for one or more sub-ranges. 

\begin{figure}
\centering 
\includegraphics[width=\linewidth]{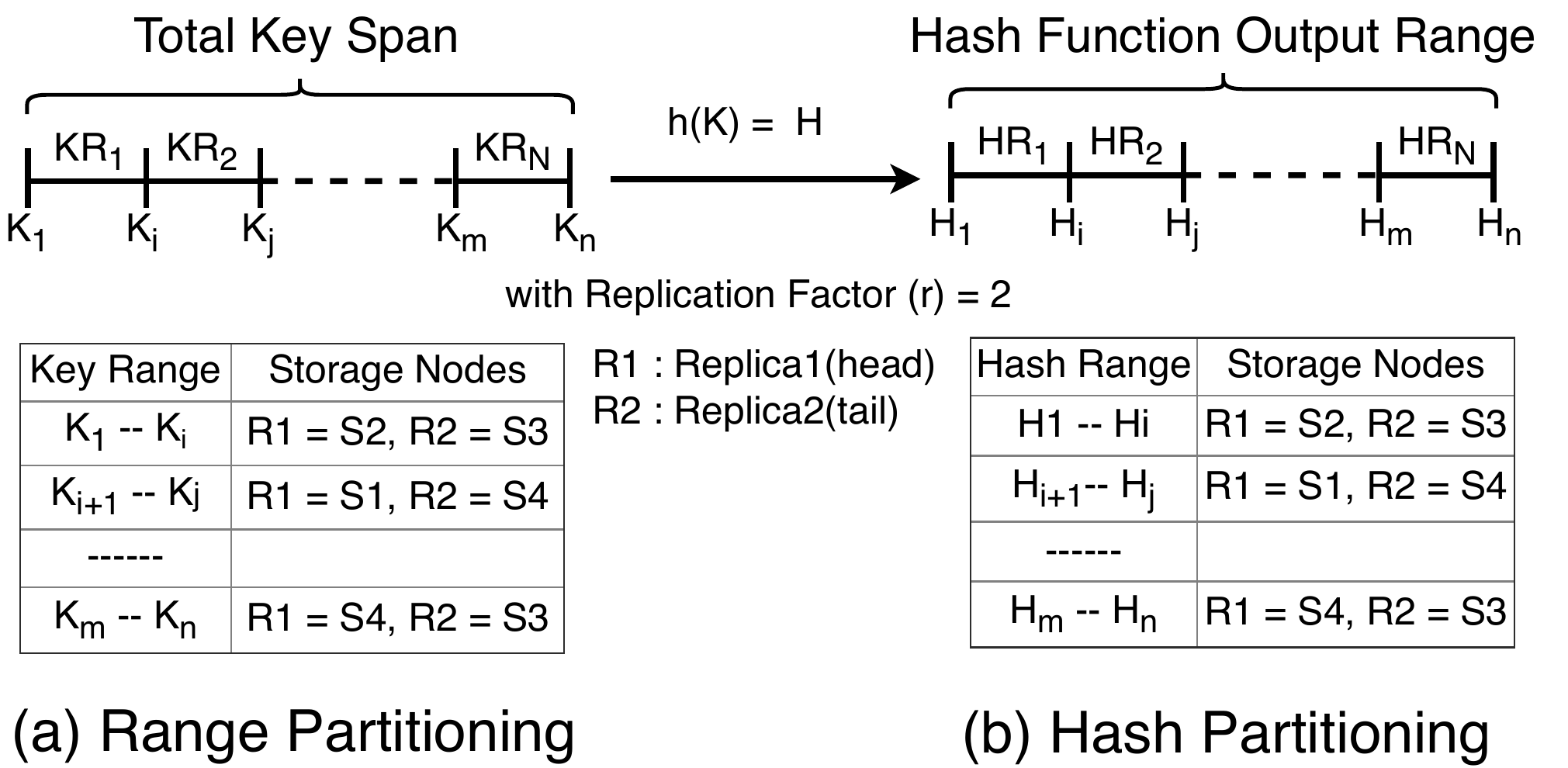}
\caption{\kvnet~Data Partitioning and Replication}
\label{fig:range_mapping_table}
\end{figure}

Each storage node has LevelDB~\cite{leveldb} or any of its enhanced versions installed where keys are stored in lexicographic order on SSTs (Sorted String Tables). This key-value library is responsible for managing the sub-ranges associated to the storage node and handling the key-value pair operations directed to the storage node. The advantage of this partitioning scheme is that range queries on keys can be supported, but unfortunately, this scheme suffers from load imbalance problem discussed in Section~\ref{sec:query_statistics}. Some sub-ranges contain popular keys which receive more requests than others. The mapping table for this type of partitioning represents the key range and the associated nodes where the data resides as shown in \kvfig\ref{fig:range_mapping_table}(a). Support of range partitioning in the switch data plane is discussed in Section~\ref{sec:data_plane_mapping}.

\vspace{10 px}
\noindent\textbf{Hash Partitioning.} In hash partitioning, each key is hashed into a 20-byte fixed-length digest using RIPEMD160~\cite{ripMD} which is an extremely random hash function, ensures that records are distributed  uniformly across the entire set of possible hash values. We developed a variation of the consistent hashing~\cite{consistent_hashing} to distribute the data over multiple storage nodes. The whole output range of the hash function is treated as a fixed space. This space is partitioned into sub-ranges, each sub-range represents a consecutive set of hashing values. These sub-ranges are distributed evenly on the storage nodes. Each storage node can be responsible for one or more sub-ranges based on its load, and each sub-range is assigned to one node (or multiple nodes depending on the replication factor). 

Like the consistent hashing, partitioning one of the sub-ranges or merging two sub-ranges due to the addition or removal of nodes affects only the nodes that hold these sub-ranges and other nodes remain unaffected. Each data item identified by a key is assigned to a storage node if the hashed value of its key falls in the sub-range of hashing values which the storage node is responsible for. This method requires a mapping table, shown in \kvfig\ref{fig:range_mapping_table}(b), where each record represents a hashed sub-range and its associated nodes where data resides. Support of hash partitioning in the switch data plane is discussed in Section~\ref{sec:data_plane_mapping}. The disadvantage of this  technique, like all hashing partitioning techniques, is that range queries can not be supported. On the storage nodes side, data is managed in hash-tables and collisions are handled using separate chaining in the form of binary search tree. 

For both partitioning techniques, we assume that there is no fixed space assigned to each sub-range (partition) on the storage node (i.e., the space that each partition consumes on a storage node can grow as long as there is available space on the storage node). Unfortunately, multiple insertions to the same partition may exceed the capacity of the storage node. In this case, The sub-range of this partition will be divided into two smaller sub-ranges. One of these small sub-ranges will be migrated to another storage node with available space. Other storage nodes that have the same divided sub-range with available space keep the data of the whole sub-range and manipulate it as it is. The mapping table will be updated with the new changes of the divided sub-ranges. 

\begin{figure}
\centering 
\includegraphics[width=\linewidth]{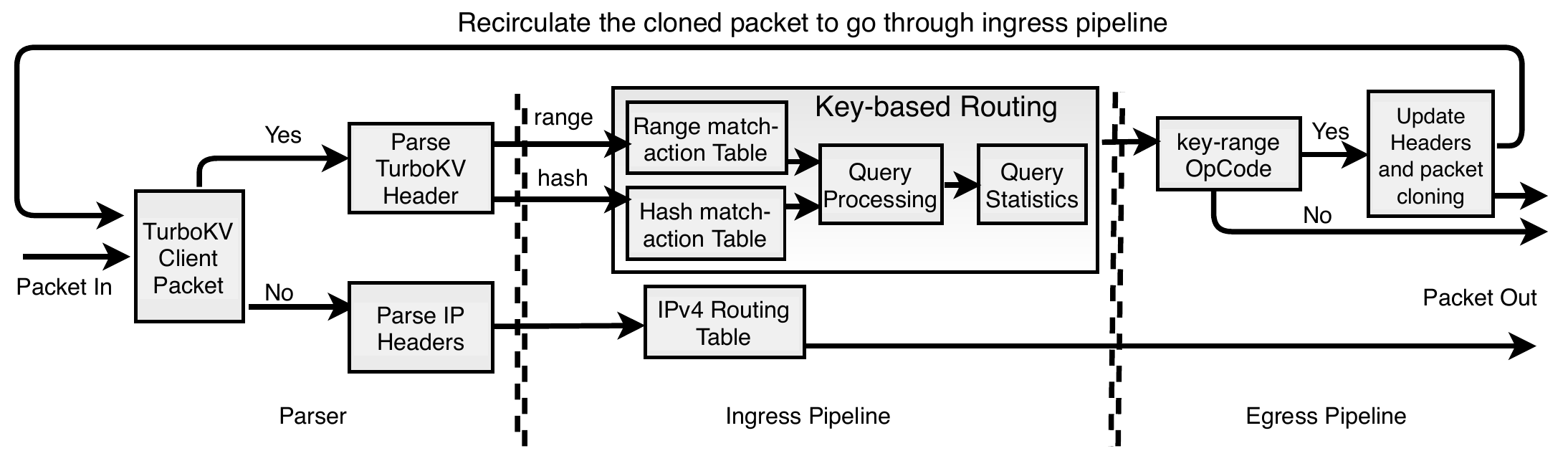}
\caption{Logical View of \kvnet~Data Plane Pipeline}
\vspace{-10 px}
\label{fig:switch_pipeline}
\end{figure} 

\subsubsection{Data Replication}

To achieve high availability and durability, data is replicated on multiple storage nodes. The data of each partition (sub-range) is replicated as one unit on different storage nodes. The replication factor ($r$) can be adjusted according to application needs. The list of nodes that is responsible for storing a particular partition is called the \textit{replica list}. This \textit{replica list} is stored in the mapping table as shown in \kvfig\ref{fig:range_mapping_table}. 

\begin{figure}
 \centering 
 \includegraphics[width=\linewidth]{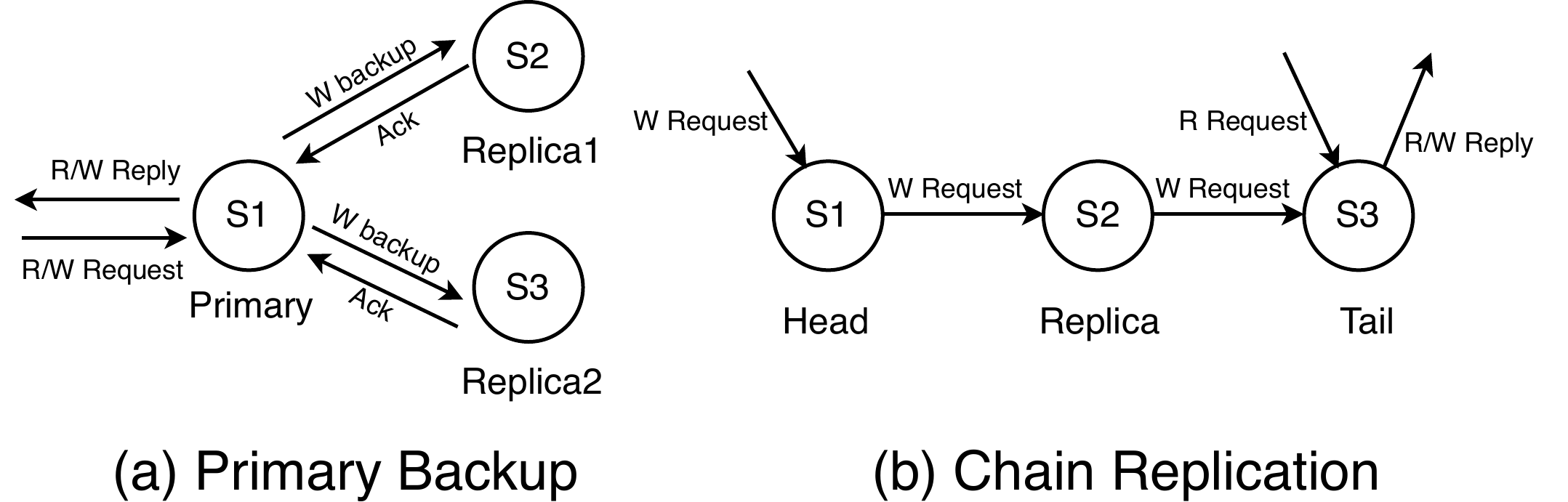}
 \caption{Replication Models}
 \vspace{-10 px}
 \label{fig:replication_model}
\end{figure}

\kvnet~follows the chain replication (CR) model~\cite{chainreplication}. Chain replication~\cite{chainreplication} is a form of primary backup protocols to control clusters of fail-stop storage servers. This approach is intended for supporting large-scale storage services that exhibit high throughput and availability without sacrificing strong consistency guarantees. In this model as shown in \kvfig\ref{fig:replication_model}(b), the storage nodes, holding replicas of data, are organized in a sequential chain structure. Read queries are handled by the tail of the chain, while write queries are sent to the head of the chain, the head process the request and forwarded it to its successor in the chain structure. This process continues till reaching the tail of the chain. Finally, the tail processes the request and replies back to the client. 

In CR, Each node in the chain needs to know only about its successor, where it forwards the request. This makes the CR simpler than the classical primary backup protocol~\cite{primarybackup}, shown in \kvfig\ref{fig:replication_model}(a) which requires the primary node to know about all replicas and keep track of all acknowledgement received from all replicas. Moreover, In chain replication, write queries also use fewer messages than the classical primary backup protocol, (n+1) instead of (2n) where n is number of nodes.  With r replicas, \kvnet~can sustain up to ($r$-1) node failures as requests will be served with other replicas on the chain structure.

\subsubsection{Management Support in Switch Data Plane}\label{sec:data_plane_mapping}

\begin{figure}
\centering 
\includegraphics[width=\linewidth]{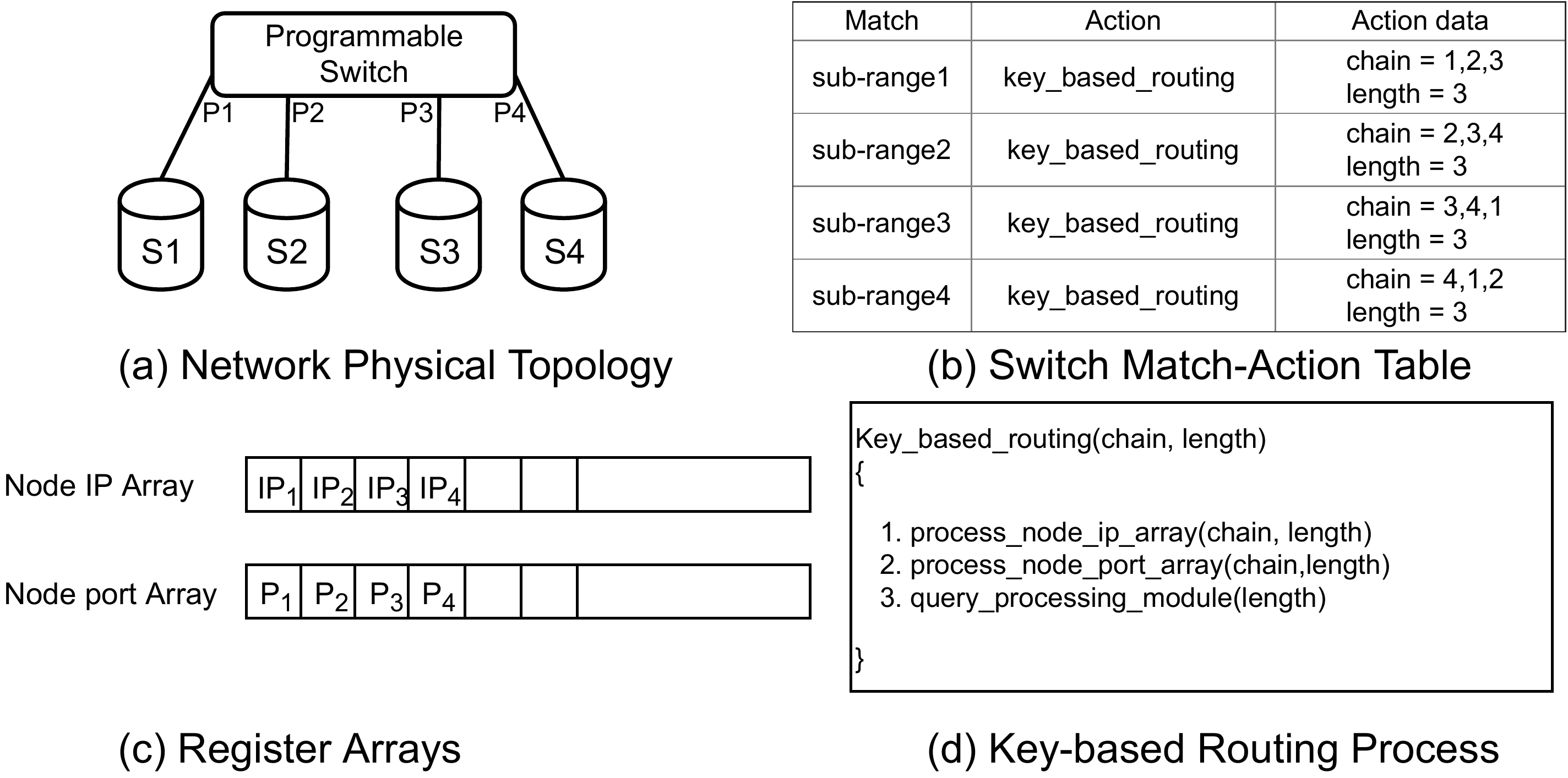}
\caption{\kvnet~Partition Management inside Switch}
\vspace{-10 px}
\label{fig:switch_data_plane}
\end{figure}

In \kvnet, the switch data plane has three types of match-action tables: range partition management table, hash partition management table and the normal IPv4 routing table. We will discuss the design of the partition management tables which are related to the key-based routing protocol for the rack scale cluster shown in \kvfig\ref{fig:switch_data_plane}(a). The design of both tables is the same but the value we used for matching and the value we match against are different based on the type of partitioning. We will refer to the value we use for matching as the \textit{matching value}, this value represents the key itself in case of range partitioning and the hashed value of the key in case of hash partitioning. 

The partition management match-action table design is shown in \kvfig\ref{fig:switch_data_plane}(b). Each record in the table consists of three parts: match, action and action data. The match represents the value that we match the \textit{matching value} against, we refer to it as a \textit{sub-range}. This \textit{sub-range} represents the start and end keys of a sub-range from the whole key span in range partitioning, or the start and end hash values of a consecutive set of hash values in hash partitioning. The action represents the key-based routing that will be executed when a \textit{matching value} falls within the \textit{sub-range}. The action data consists of two parts: chain and length. Chain represents the forwarding information for nodes forming the chain of the sub-range. This information includes node's IP address and the port from the switch to the storage node. Nodes' information is sorted according to node's position in the chain structure (i.e., first node is the head and last node is the tail), and is used in updating packet during the action execution.

In \kvnet, each node holds the data of one or more sub-ranges. This makes each node's forwarding information appears more than once in the match-action table records. \kvnet~uses two arrays of registers in the switch's data plane to save the forwarding information: node IP array, and node port array. For each storage node, the forwarding information is stored at the same index in the two arrays as shown in \kvfig\ref{fig:switch_data_plane}(c). For example, the information of storage node \textit{S1} is stored at index 1 in the two arrays and the same for the other storage nodes. The index of the storage nodes in the register arrays is stored as action data in the match-action table records to form the chain as shown in \kvfig\ref{fig:switch_data_plane}(b). The key-based routing uses these indexes to process the register arrays and fetch the forwarding information for the replica nodes. This information is used by the query processing module to forward packets to their next hop.

\subsection{Network Protocol Design}\label{sec:network_design}

\begin{figure}
\centering 
\includegraphics[width=\linewidth]{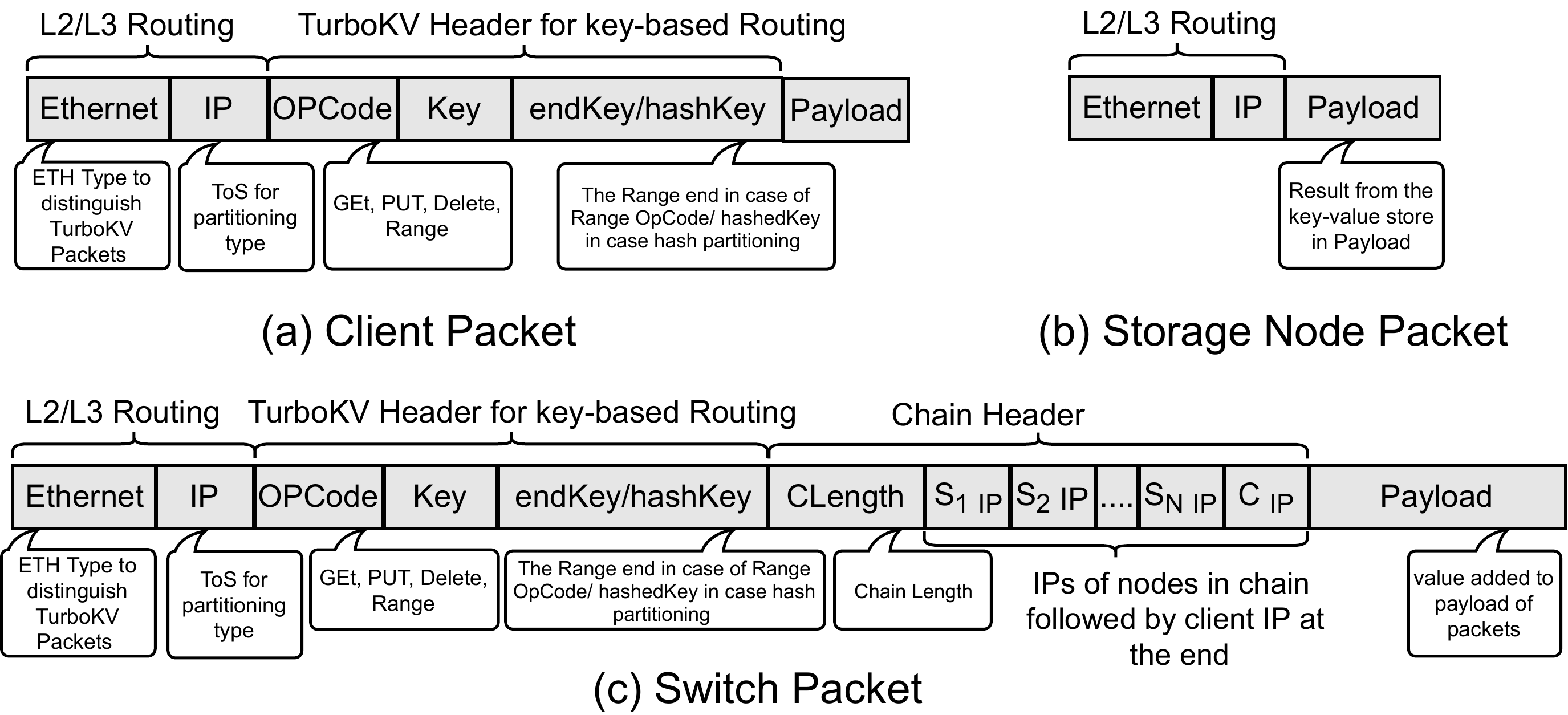}
\caption{\kvnet~Packet Format}
\vspace{-15 px}
\label{fig:packet_format}
\end{figure}

\textbf{Packet Format.} 
Figure~\ref{fig:packet_format}(a) shows the format of \kvnet~request packet sent from clients. The programmable switches use the \texttt{Ethernet Type} in the Ethernet header to identify \kvnet~packets, and execute the key-based routing. Other switches in the network do not need to understand the format of \kvnet~header, and treat all packets as normal IP packets. The \texttt{ToS} (Type of Service) in the IP header is used to distinguish between three types of \kvnet~packets: range partitioned data packet, hash partitioned data packet, and \kvnet~packet previously processed by the switch. The \kvnet~header consists of three main fields: \texttt{OpCode}, \texttt{Key}, and \texttt{endKey/hashedKey}. The \texttt{OpCode} is a one-byte field which stands for the code of the key-value operation (\textit{Get, Put, Del, and Range}). \texttt{Key} stores the key of a key-value pair. In \textit{Range} operation, \texttt{Key} and \texttt{endKey/hashedKey} are used to represent the start and end of the key range, respectively. In case of hash partitioning, the \texttt{endKey/hashedKey} is set with the hashed value of the key to perform the routing based on it instead of the key itself.

After the client packet is processed by the programmable switch, the switch adds the chain header shown in \kvfig\ref{fig:packet_format}(c). This header is used by the storage nodes for the chain replication model. It includes two fields. The first field is the number of nodes which the packet passes by in the chain including the client IP (\texttt{CLength}). The second field has these nodes' IP addresses ordered according to their position in the chain followed by the client IP at the end. The packet format of the reply from storage node to client is a standard IP packet shown in Figure~\ref{fig:packet_format}(b). The result is added to the packet payload.

\vspace{10 px}
\noindent\textbf{Network Routing.}
\kvnet~uses a key-based routing protocol to route packets from clients to storage nodes. The client sends the packet to the network. Once it reaches the programmable switch, the switch extracts the \texttt{Key} (in case of range partitioning) or the \texttt{endkey/hashedKey} (in case of hash partitioning) from \kvnet~header, and looks up the corresponding key-based match-action table using the value of the extracted field (\textit{matching value}). If there is a hit, the switch fetches the chain nodes' information from the registers based on the specified indexes in the match-action table. Then, the switch processes this information based on the query type (\texttt{OpCode}). After that, the switch updates the packet with the target node's information, and forwards it to the next hop on the path to this target node.

The switch uses the \textit{range matching} for table lookup, in which, it matches the \textit{matching value} against the sub-range in the corresponding key-based match-action table. If the \textit{matching value} falls within one of the sub-ranges (hit), the key-based routing action is processed using the action data. The programmable switches route the previously processed \kvnet~packets and the storage node to client packets using the standard L2/L3 protocol without passing the key-based routing and perform the match-action based on the destination IP in the IP header.

\subsection{Key-value Storage Operations}\label{sec:query_processing}

\begin{figure}
\centering 
\includegraphics[width=\linewidth]{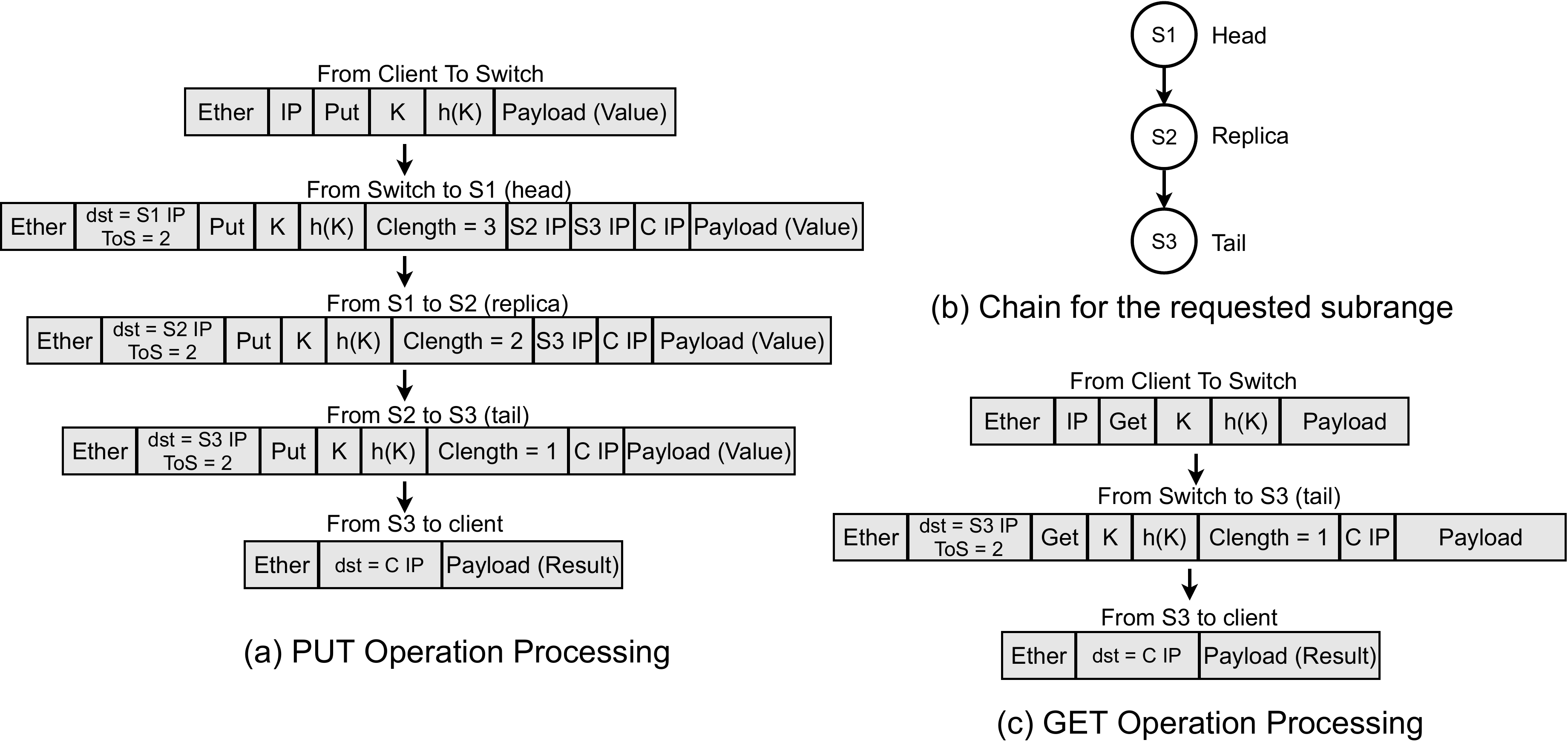}
\caption{\kvnet~KV Storage Operations}
\vspace{-10 px}
\label{fig:query_processing}
\end{figure} 

\noindent\textbf{PUT and DELETE Queries.}
In chain replication, PUT and DELETE queries are processed by each node along the chain starting from the head and replied by the tail. On the switch side, after processing one of the key-based match-action tables with the \textit{matching value} and fetching the corresponding chain information from the register arrays, the query processing module sets the destination IP address in the IP header with the IP address of the chain head node and changes the \texttt{ToS} value to mark the packet as previously processed. The egress port is set with the forwarding port of the chain head node, then the chain header is added to the packet with \texttt{CLength} equals to the length of the chain and the chain nodes' IP addresses ordered according to their position in the chain followed by the client IP at the end as shown in \kvfig\ref{fig:query_processing}(a). Finally, the packet is forwarded to the head of chain node. 

As shown in \kvfig\ref{fig:query_processing}(a), when the packet arrives at a storage node, it is processed by \kvnet~storage library. The node updates its local copy of the data. Then, it reads the chain header, sets the destination address in the IP header with the IP of its successor and reduces the \texttt{CLength} by 1, then forwards the packet to its successor. Packets received by the tail node, with \texttt{CLength} = 1, have their chain header and \kvnet~header removed, and the result is sent back using the client IP as the destination address in the IP header.  

\vspace{10 px}
\noindent\textbf{GET Queries.}
Following the chain replication, GET queries are handled by the tail of the chain. After performing the matching on the \textit{matching value} and fetching the corresponding chain information from the register arrays, the query processing module sets the destination IP address in the IP header with the IP address of the chain tail node and changes the \texttt{ToS} value to mark the packet as previously processed. The egress port is set with the forwarding port of the chain tail node, then the chain header is added to the packet with \texttt{CLength} equals to 1 and one node IP which represents the client IP as shown in \kvfig\ref{fig:query_processing}(c). Finally, the packet is forwarded to the storage node. When the packet arrives at the storage node, it is processed by \kvnet~storage library. The query result is added to the payload of the packet. and the client IP is popped up from the chain header and put in the destination address in the IP header. The chain and \kvnet~headers are removed from the packet and the result is sent back to the client.  

\vspace{10 px}
\noindent\textbf{Range Queries.}
If the data is range partitioned, \kvnet~can handle range queries. The requested range in the \kvnet~header may span multiple storage nodes. Thus, the switch divides the range into several sub-ranges, each sub-range corresponds to a packet. Each of these packets contains the start and end keys of the corresponding sub-range. Each packet is handled by the switch like a separate read query and forwarded to the tail node of its partition's chain.  Unfortunately, the switch cannot construct new packets from scratch on its own. Therefore, in order to achieve the previous scenario, we placed the range operation check in the egress pipeline as shown in \kvfig\ref{fig:switch_pipeline}. We use the \texttt{clone} and \texttt{circulate} operations, supported in the architecture of the programmable switches and P4, to solve the packet construction problem. When a range operation is detected in the egress pipeline, the packet is processed as shown in Algorithm~\ref{range_algorithm}.

\begin{figure}
\makebox[\linewidth]{%
\begin{minipage}{\dimexpr\linewidth}
\begin{algorithm}[H]
\caption{Range Query Handling}\label{range_algorithm}
\fontsize{6}{6}\selectfont
\begin{algorithmic}[1]
	\STATE \textbf{Input pkt:} packet entering the egress pipeline
	\STATE \textbf{matched\_subrange:} the subrange where the start key of the requested range falls 
	\STATE \textbf{Output pkt\_out:} packet to be forwarded to the next hop
	\STATE \textbf{Output pkt\_cir:} packet to be circulated and sent to ingress pipeline as new packet
	\STATE \emph{Begin}:
		\STATE pkt\_out = pkt~~~~~// clone the packet
		\IF{pkt.OpCode == range}
			\STATE // check if range spans multiple nodes
			\IF{pkt.request.endKey > matched\_subrange.endKey} 
				\STATE pkt\_cir = pkt~~~~~// clone the packet			
				\STATE pkt\_out.request.endKey = matched\_subrange.endKey
				\STATE pkt\_cir.request.Key = Next(matched\_subrange.endKey)
	 		\ENDIF
		\ENDIF
		\IF{pkt\_cir.exist()}
			\STATE circulate(pkt\_cir)~~~~~// send packet to ingress pipeline again
		\ENDIF		
		\STATE send\_to\_output\_port(pkt\_out)
\end{algorithmic}
\end{algorithm}
\end{minipage}}
\end{figure}

\section{\kvnet~Control Plane Design}\label{sec:control_design}

The control plane provides load balancing module based on the query statistics collected in the data plane. It also provide a failure handling module to guarantee availability and fault tolerance. In this section, we discuss how the control plane supports these functions.

\begin{figure}
 \centering 
 \captionsetup{justification=centering}
 \includegraphics[width=\linewidth]{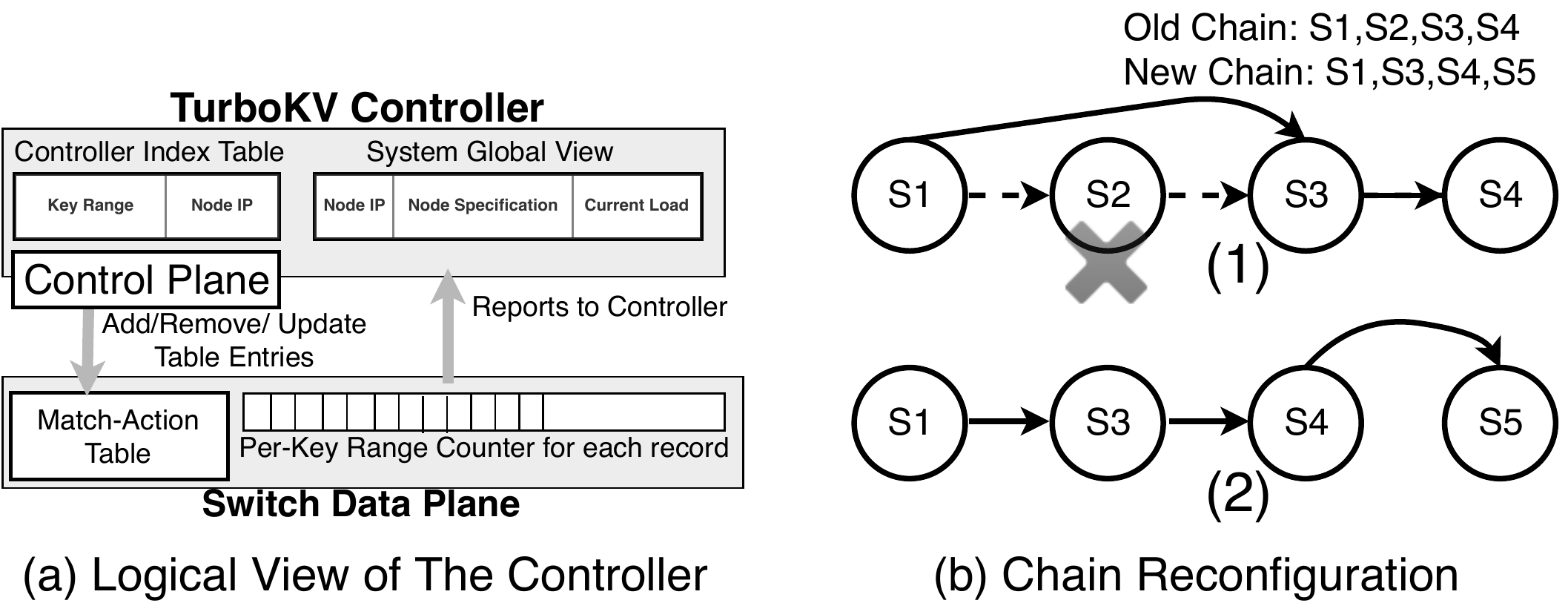}
 \caption{\kvnet~Control Plane}
 \vspace{-10 px}
 \label{fig:query_statistics}
\end{figure}

\subsection{Query Statistics and Load Balancing}\label{sec:query_statistics}

In \kvnet, the data plane has a query statistics module to provide query statistics reports to the \kvnet~controller. Thus, the controller can estimate the load of each storage node, and make decisions to migrate part of the popular data to one of the under-utilized storage nodes. As shown in \kvfig\ref{fig:query_statistics}(a), the switch's data plane maintains a per-key range counter for each key range in the match-action table. Upon each hit for a key range, its corresponding counter is incremented by one. The controller receives reports periodically from the data plane including these statistics, and resets these counters in the beginning of each time period. Then, the controller  compares the received statistics with the specifications of the storage nodes. If a storage node is over-utilized, the controller migrates a subset of the hot data in a sub-range to another storage node and reconfigures the chain of the updated sub-range. Then, the controller updates records in the match-action table of the switches with the new chain configurations. After the sub-range's data is migrated to other storage nodes, the old copy is removed from the over-utilized one. Currently, the controller follows a greedy selection algorithm to select the least utilized node where data will be migrated.      

\kvnet~uses the physical migration of the data to achieve load balancing between the storage nodes. This approach adapts with all kinds of workloads compared to the caching approach. In caching\cite{netcache}, the cache absorbs the read requests of the very popular key-value pairs, which makes it performs well in highly skewed read-only workload, but the effect of caching in load balancing decreases if the workload's ratio of updates increases. This behavior resulted from the invalid cached key-value pairs, which make the request directed to the target storage node to retrieve the valid pairs. 

\subsection{Failures Handling}\label{sec:failure_handling}

We assume that the controller process is a reliable process and it is different from the SDN controller. We also assume that links between storage nodes and switches in data centers are reliable and are not prone to failures.

\vspace{10 px}
\noindent\textbf{Storage Node Failure.}
When the controller detects a storage node failure, it reconfigures the chains of the sub-ranges on the failed storage node and updates their corresponding records in the key-based match-action table through the control plane. The controller removes the failed storage node from its position in all chains. In each chain, the predecessor of the failed node will be followed by the successor of the failed node, reducing the chain length by 1 as shown in \kvfig\ref{fig:query_statistics}(b). If the failed node was the head of the chain, then the new head will be its successor. If the failed node was the tail of the chain, then the new tail will be its predecessor.
Reducing the chain length by 1 makes the system able to sustain less number of failures. That is why the controller distributes the data of the failed node in sub-range units among other functional nodes, and adds these new nodes at the end of these sub-ranges' chains in their corresponding records in the key-based match-action table. This process of sub-range redistribution restores the chain to its original length. 

\vspace{10 px}
\noindent\textbf{Switch failure.} The storage servers in the rack  of the failed switch would lose access to the network. The controller will detect that these storage servers are unreachable. The controller treats these storage servers as failed storage nodes, and distributes the load of these storage servers among other reachable servers as described before. Then, the failed switch needs to be rebooted or replaced. The new switch starts with an index table which contains all the key ranges handled by its connected storage servers.

\section{Scaling Up to Multiple Racks} \label{sec:scaling_up}

\begin{figure}
 \centering 
 \includegraphics[width=0.45\linewidth]{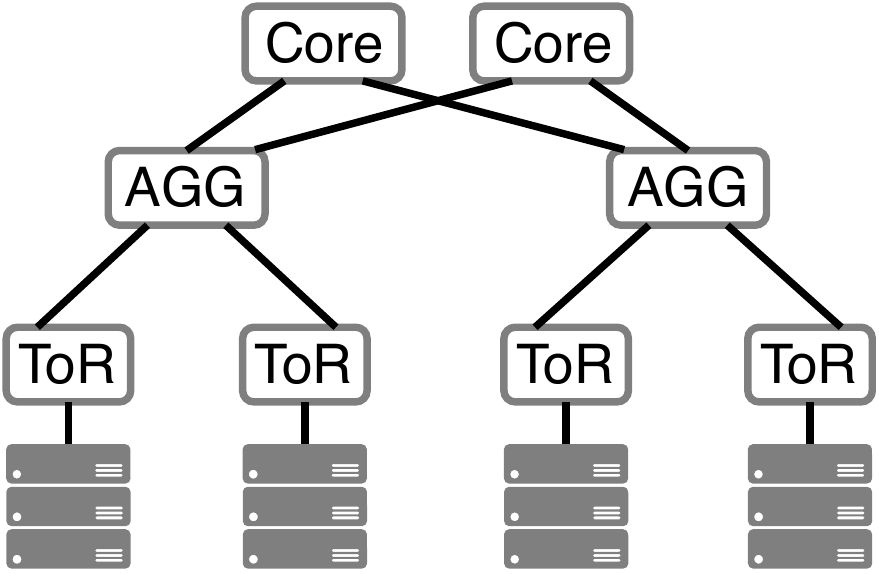}
 \caption{Scaling Up inside Data center Network}
 \vspace{-10 px}
 \label{fig:scaling_up}
\end{figure}

We have discussed the in-switch coordination for distributed key-value stores within a rack of storage nodes with the Top-of-Rack (ToR) switch as the coordinator. We now discuss how to scale out the in-switch coordination in the data center network.
\kvfig\ref{fig:scaling_up} shows the network architecture of data centers. All the servers in the same rack are connected by a ToR switch. In the highest level, there are Aggregate switches (AGG) and Core switches (Core). 

To scale out distributed key-value stores with in-switch coordination, we develop a "hierarchical indexing" scheme. Each ToR switch has the directory information of all sub-ranges located on its connected storage nodes as described before in \kvfig\ref{fig:switch_data_plane}. In addition to the IPv4 routing table, each AGG switch has two range match-action tables (range and hash), where each table consists of the sub-ranges in its connected ToR switches. The Core switches have the range match-action tables of sub-ranges in its connected AGG switches. With each sub-range in either the AGG or Core switches, the action data represents only the forwarding port towards the head or the tail of the sub-range's chain. No chains are stored in these switches. When a packet is received by an AGG switch or core switch, this packet is processed by the key-based routing protocol without adding any chain header to the packet and forwarded to one of the ports towards the head or tail of the sub-range chain based on the query (write or read respectively). When the packet arrives at ToR switch, the switch processes the packet as discussed in Section~\ref{sec:query_processing}. Replicas of a specific sub-range may be located on different racks. This design leverages the existing data center network architecture and does not introduce additional network topology changes.

\section{Implementation Details}\label{sec:implementation}

We have implemented a prototype of \kvnet, including all switch data plane and control plane features, described in Section~\ref{sec:data_design} and Section~\ref{sec:control_design}. We have also implemented the client and server libraries that interface with applications and storage nodes, respectively, as described in Section~\ref{sec:architecture}.
The switch data plane is written in P4 and is compiled to the simple software switch BMV2~\cite{bmv2} running on Mininet~\cite{mininet}. The key size of the key-value pair is 16 bytes with total key range spans from 0 to $2^{128}$. This range is divided into index records which are saved on the switch data plane. We used 4 register arrays, one for saving the storage nodes' IP addresses, one for saving the forwarding port of the storage nodes, one for counting the read access requests of the indexing records and the last one for counting the update access requests of the indexing records. The controller is able to update/read the values of these registers through the control plane. It also can add or remove table entries to balance the load of the storage nodes. The switch data plane does not store any key-value pairs as these pairs are saved on the storage nodes. This approach makes \kvnet~consumes a small amount of the on-chip memory leaving enough space for processing other network operations. The controller is written in Python and can update the switch data plane through the switch driver by using the Thrift API generated by the P4 compiler.

\begin{figure}
 \centering 
 \includegraphics[width=\linewidth, height=150 px]{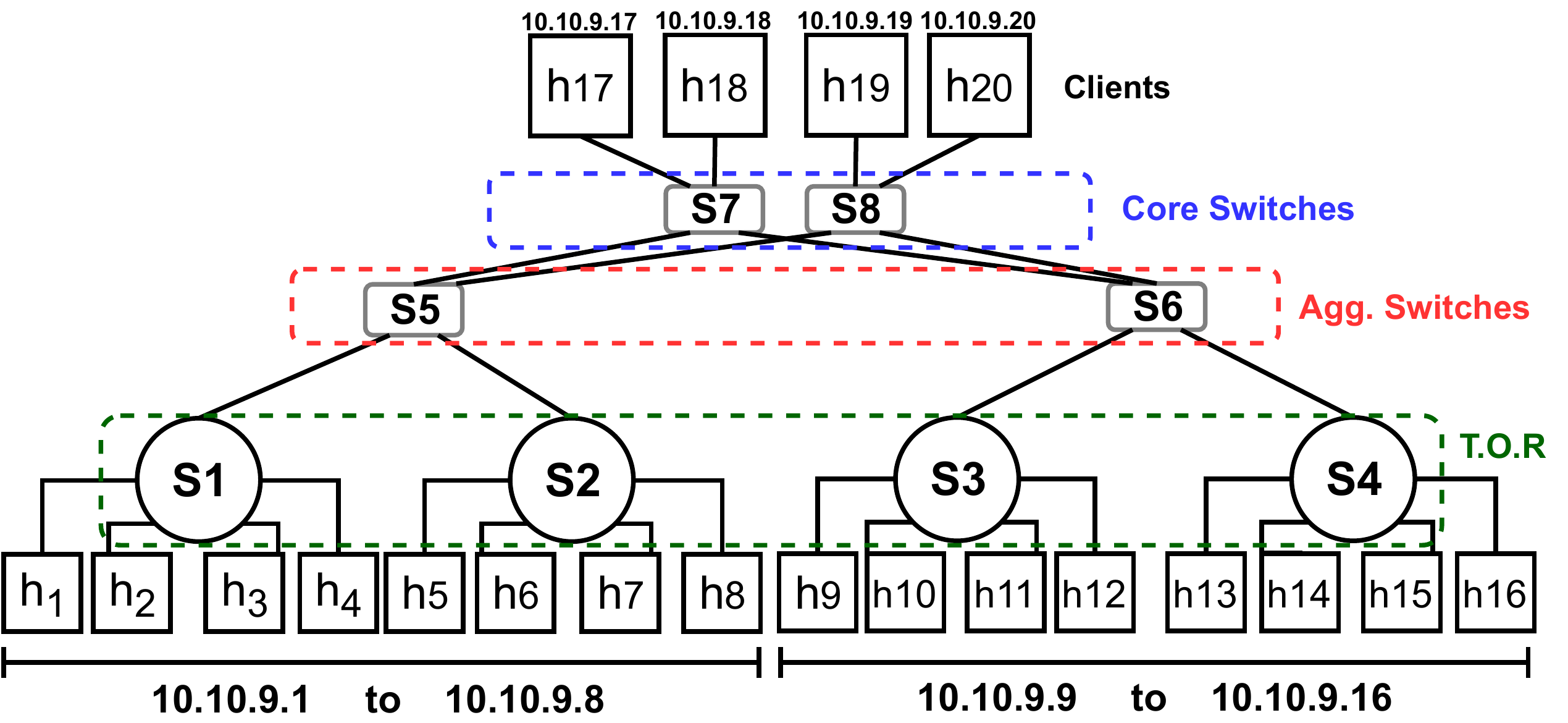}
 \caption{Experiment Topology}
 \vspace{-10 px}
 \label{fig:topology}
\end{figure}

The client and server libraries are written in Python using Scapy~\cite{scapy} for packet manipulation. The client can translate a generated YCSB~\cite{ycsb} workload with different distributions and mixed key-value operations into \kvnet~packets' format and send them through the network. We used Plyvel~\cite{plyvel} which is a Python interface for levelDB~\cite{leveldb} as the storage agent. The server library translates \kvnet~packets into Plyvel format and connects to levelDB to perform the key-value pair operations. We used chain replication with chain length equals to 3 for data reliability.  

\section{Performance Evaluation}\label{sec:results}

In this section, we provide the experimental evaluation results of \kvnet~running on the programmable switches. These results show the performance improvement of \kvnet~on the key-value operations latency and system throughput.

\begin{figure*}
\centering 
\subfigure[Throughput vs Skewness - Read only]{\label{fig:throughput_readonly}\includegraphics[width=0.3\linewidth]{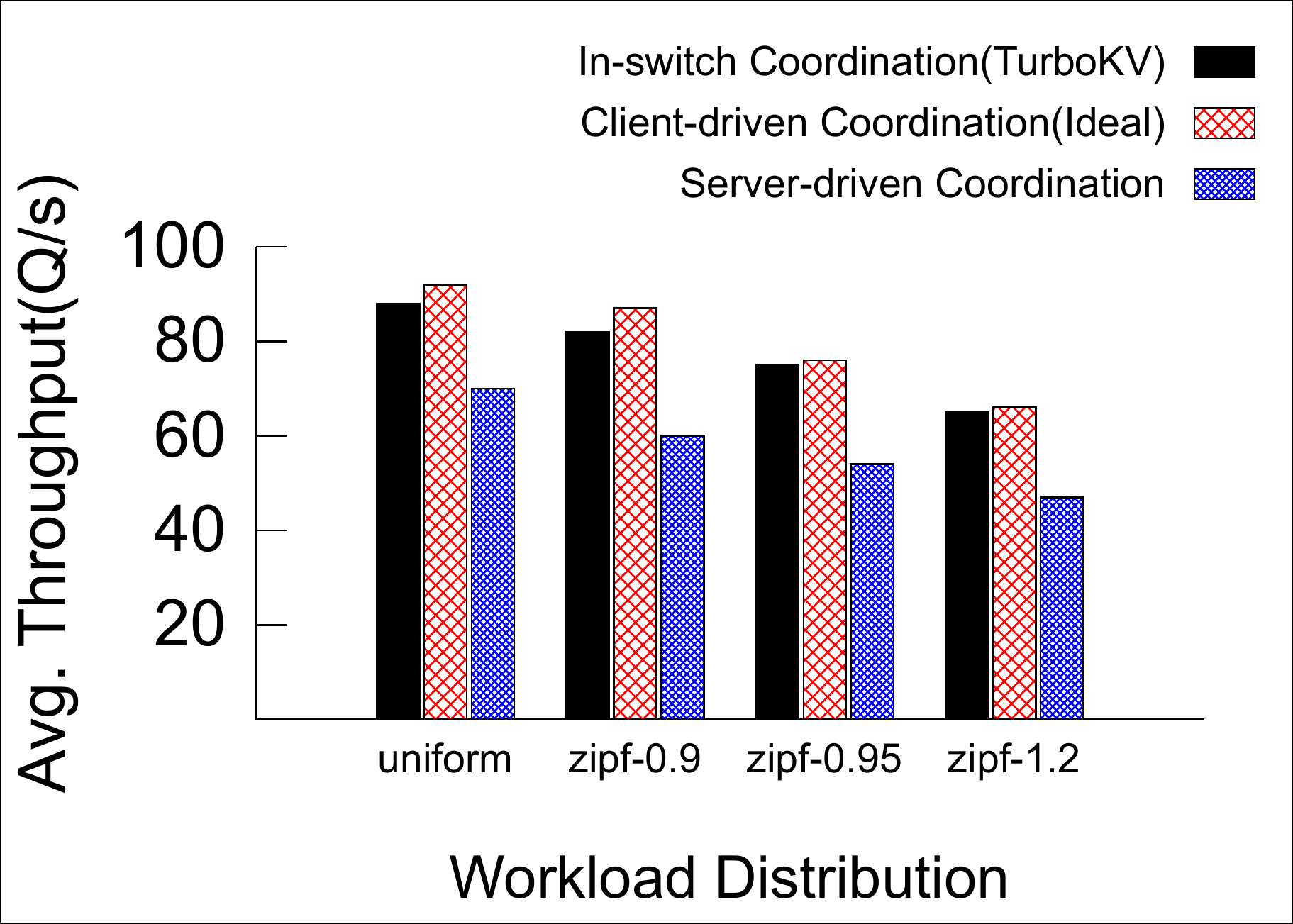}}
\hspace{0.03\linewidth}
\subfigure[Throughput vs Write Ratio - Uniform]{\label{fig:write_ratio_uniform}\includegraphics[width=0.3\linewidth]{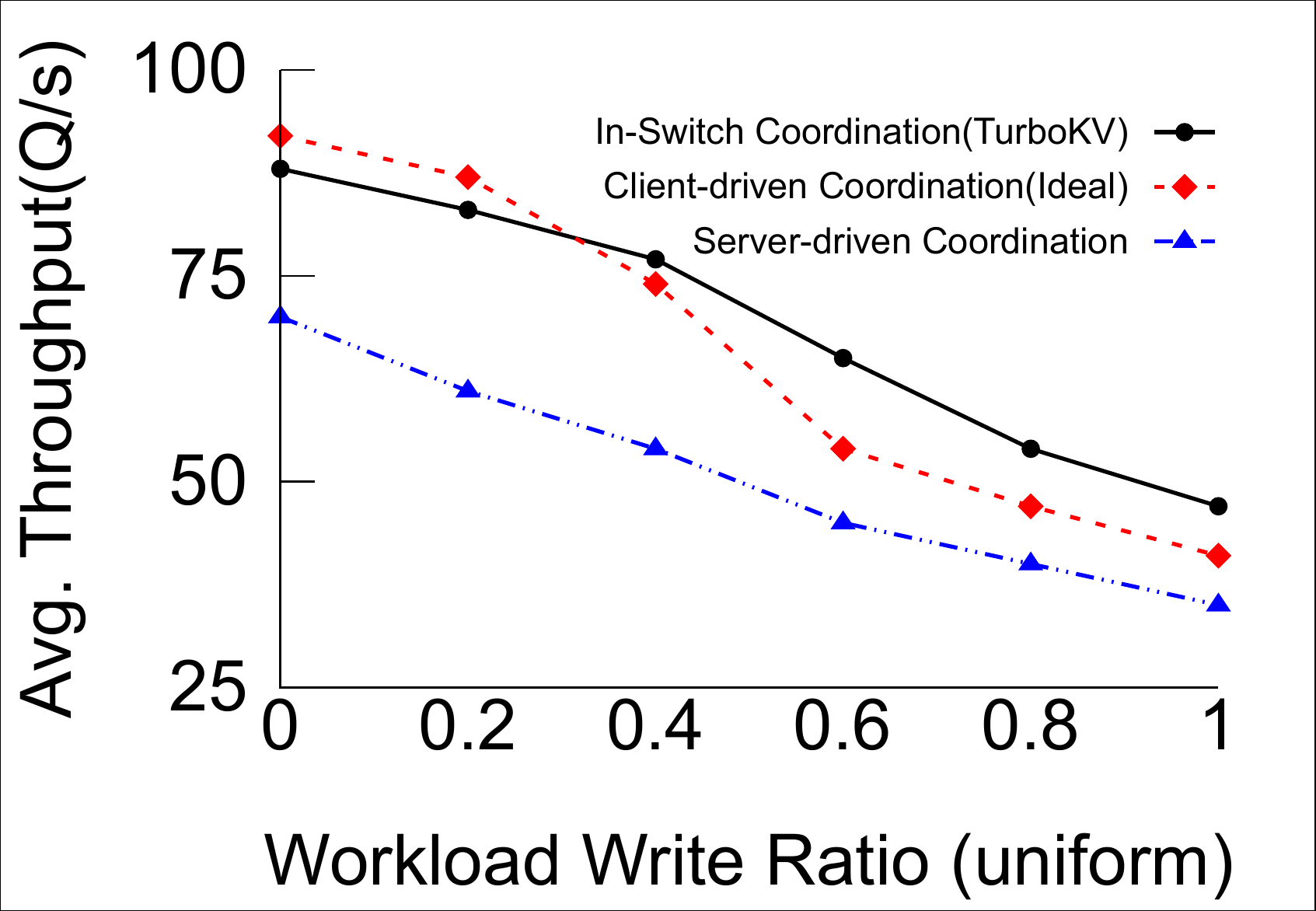}}
\hspace{0.03\linewidth}
\subfigure[Throughput vs Write Ratio - zipf0.95]{\label{fig:write_ratio_zipf}\includegraphics[width=0.3\linewidth]{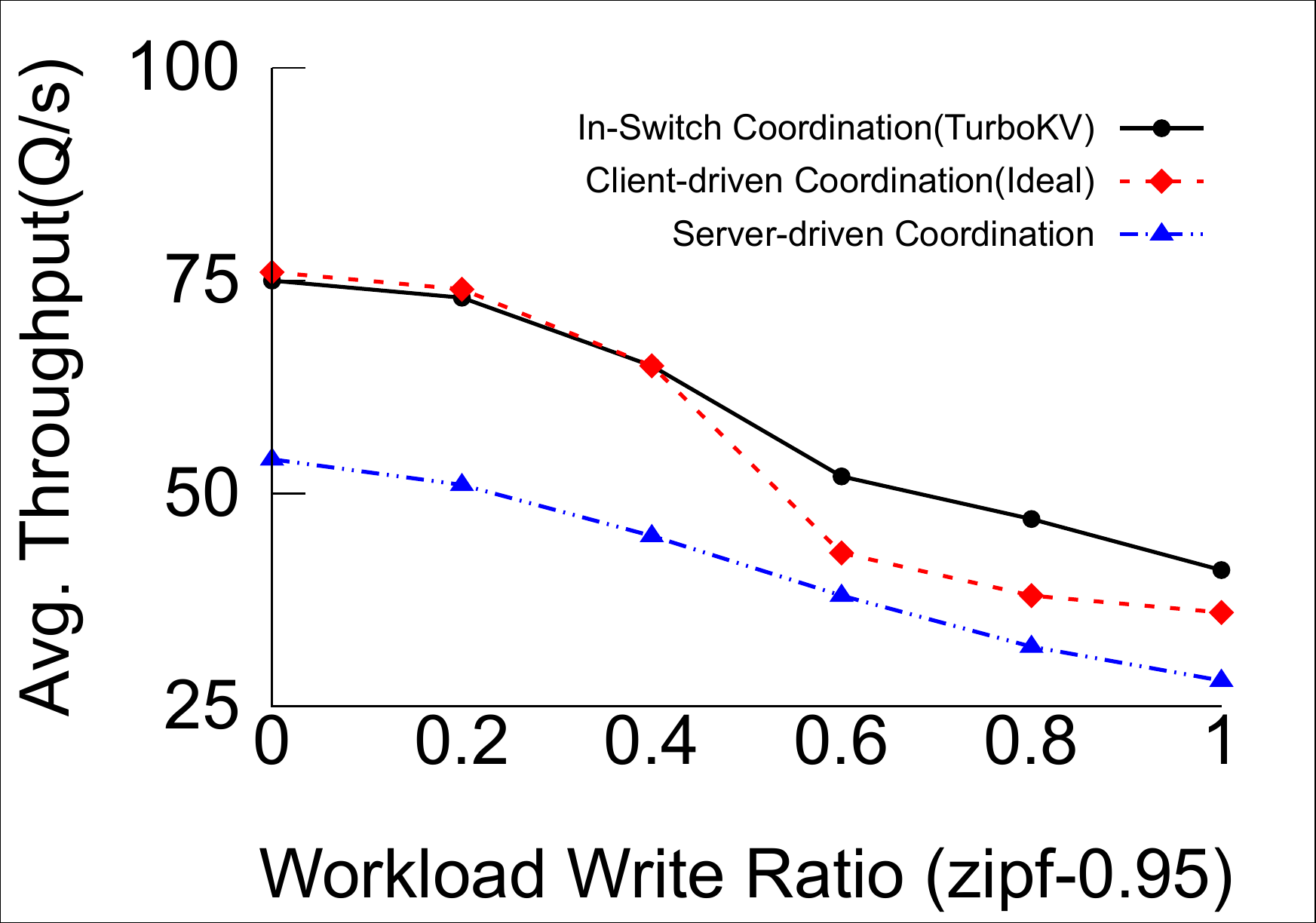}}
\caption{Effect on System Throughput}
\label{fig:effect_throughput}
\end{figure*}

\vspace{10 px}
\noindent\textbf{Experimental Setup.} Our experiments consist of eight simple software switches BMV2~\cite{bmv2} connecting 16  storage nodes and 4 clients as described in \kvfig\ref{fig:topology}. Each of the clients ($h_{17}, h_{18}, h_{19}, h_{20}$) runs the client library and generates the key-value queries. These clients represent the request aggregation servers who are the direct clients of the storage nodes in data centers. Each storage node ($h_1$, $h_2$,...., $h_{15}$, and $h_{16}$) runs the server library and uses LevelDB as the storage agent. The whole topology is running on Mininet. The data is distributed over the storage nodes using the range partitioning described in Section~\ref{sec:data_partitioning} with 128 records index table. Each storage node is responsible for 24 sub-ranges (head of chain of 8 sub-ranges, replica for 8 sub-ranges and tail of chain for 8 sub-ranges ), and stores the data whose keys fall in these sub-ranges.  

\vspace{10 px}
\noindent\textbf{Comparison.} We compared our in-switch coordination (\kvnet) with server-driven coordination and client-driven coordination described in Section~\ref{sec:introduction}. In \kvnet, the directory information is stored in the switch data plane and updated by \kvnet~controller, also the key-based routing is used to route the query from client to target storage node. In server-driven coordination which is implemented on most of existing key-value stores~\cite{dynamo,cassendra}, all the storage nodes store the directory information and can act as the request coordinator. 
In client-driven coordination, the client acts as the request coordinator. The client has to download the directory information periodically from a random storage node, because, with lots of outdated directory information, the client-driven coordination tends to act as the server-driven coordination as the wrong storage node that the client contacts will forward the request again to the target storage node. Both of client-driven and server-driven approaches route the query using the standard L2/L3 routing protocols.

Note that in our experiments, we compare with the ideal case of the client driven coordination where the client has the updated directory information and sends the query directly to the target storage node, ignoring the latency introduced by pulling this information periodically from a random storage node because this latency will depend on the client location and also the load of the storage node that the client contacts to pull this information. The ideal case of the client-driven coordination represents \textit{the least latency} that the client's request can achieve because it represents the direct path from the client to the target storage node ignoring any latency resulted from having outdated directory information which may cause extra forwarding steps in the path from the client to the target storage node. With lots of updates to the directory information of the key-value store, the ideal client-driven coordination can not be achieved in real life systems. 

\vspace{10 px}
\noindent\textbf{Workloads.} We use both uniform and skewed workloads to measure the performance of \kvnet~under different workloads. The skewed workloads follow Zipf distribution with different skewness parameters (0.9, 0.95, 1.2). These workloads are generated using YCSB~\cite{ycsb} basic database with 16 byte key size and 128 byte value size. The generated data is stored into records' files and queries' files, then parsed by the client library to convert them into \kvnet~packet format. We generate different types of workloads: read-only workload, scan-only workload, write-only workload and mixed workload with multiple write ratios.

\subsection{Effect on System Throughput}

\noindent\textbf{Impact of Read-only Workloads.} \kvfig\ref{fig:throughput_readonly} shows the system throughput under different skewness parameters with read-only queries. We compare \kvnet~throughput vs the ideal client-driven throughput and server-driven throughput. As shown in \kvfig\ref{fig:throughput_readonly}, \kvnet~performs nearly the same as the ideal client-driven coordination in the highly skewed workload (zipf-0.99 and zipf-1.2), and less than the ideal client-driven coordination by maximum of 5\% in the uniform and zipf-0.95 workloads. This result is because the in-switch coordination manages all directory information in the switch data plane and uses the key-based routing to deliver the requests to the storage nodes directly. Moreover, the in-switch coordination eliminates the load of downloading the updated directory information periodically from storage nodes, as this part is managed by the controller who will update the directory information in the switch data plane through the control plane. In addition, the in-switch coordination outperforms the server-driven coordination and improves the system throughput with minimum of 26\% and maximum of 39\%. This result is because the in-switch coordination eliminates the overhead of the load balancer and skips a potential forwarding step introduced in the server-driven coordination when a request is assigned to a random storage node.   

\vspace{10 px}
\noindent\textbf{Impact of Write Ratio.} \kvfig\ref{fig:write_ratio_uniform} and \kvfig\ref{fig:write_ratio_zipf} show the system throughput under uniform and skewed workload, respectively, with varying the workload write ratio. As shown in \kvfig\ref{fig:write_ratio_uniform} and \kvfig\ref{fig:write_ratio_zipf}, the throughput decreases as the write ratio increases for the three approaches, because each write query has to update all the copies of the key-value pair following the chain replication approach before returning the reply to the client. \kvnet~performs roughly the same as the ideal client-driven coordination in the workloads with low write ratio, but \kvnet~outperforms the client-driven coordination as the write ratio increases. This behavior is because of the chain replication implemented in the system for availability and fault tolerance. In in-switch coordination, the switch inserts all the chain nodes in the packet. When the packet arrives at a storage node, the node updates its local copy and forwards the request to the next storage node directly without any further mapping to know its chain successor. But, in the client-driven coordination, the client sends the write query to the head of chain's node. When the query arrives at the storage node, the node updates its local copy and then accesses its saved directory information to know its chain successor, then forwards the packet to it. So, when the write ratio increases, this scenario is performed for larger portion of queries which affects the system throughput. Also, \kvnet~outperforms the server-driven coordination by minimum of 26\% and maximum of 44\% in case of uniform workload, and by minimum of 37\% and maximum of 47\% in case of the skewed workload. This improvement is because of the elimination of the forwarding step when the request is assigned to a random storage node and also the elimination of further mapping steps on each storage node for knowing the chain successor.   

\subsection{Effect on Key-value Operations Latency}

\begin{figure*}
\centering 
\subfigure[Read Latency]{\label{fig:read_uniform}\includegraphics[width=0.3\linewidth]{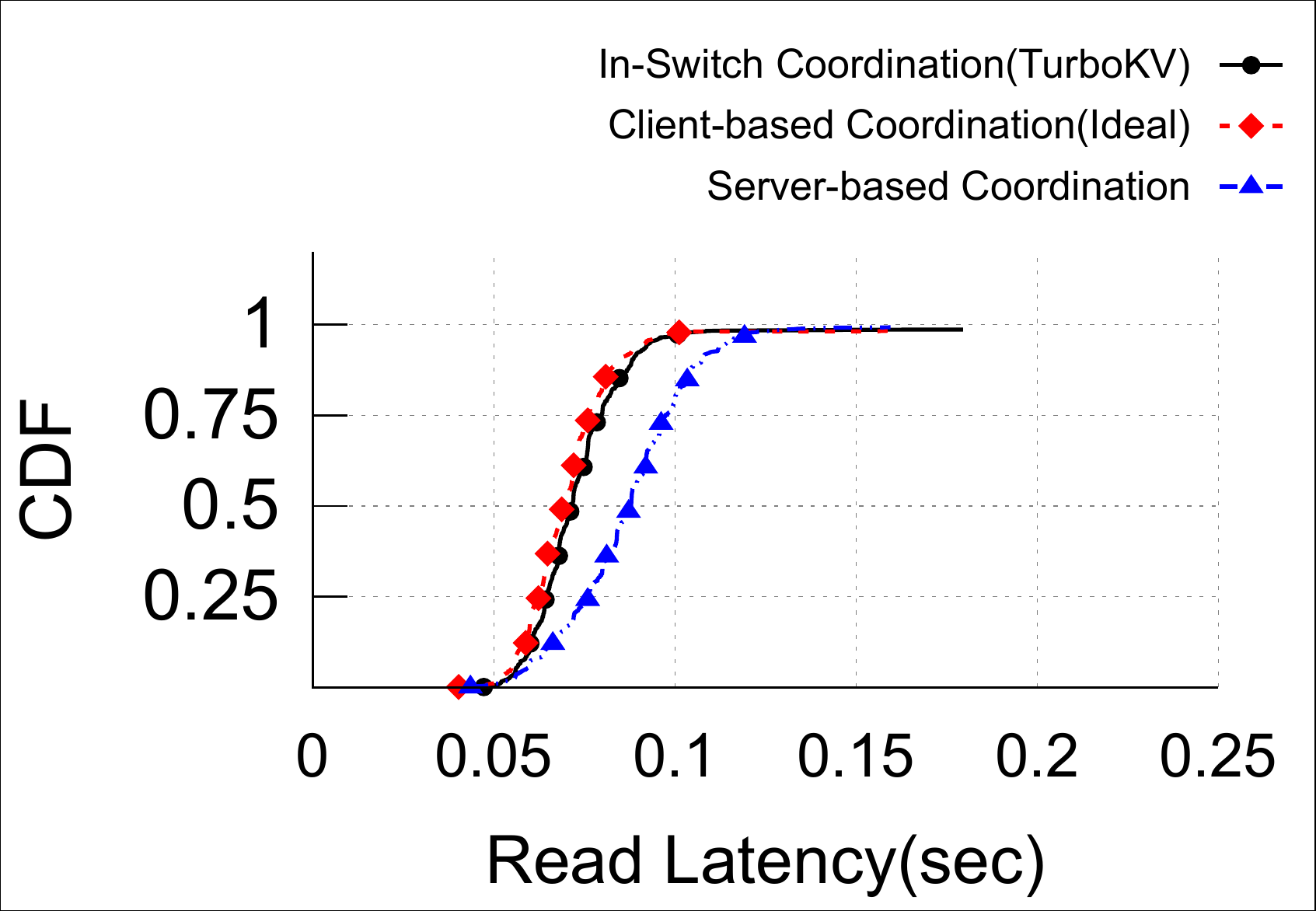}}
\hspace{0.03\linewidth}
\subfigure[Write Latency]{\label{fig:write_uniform}\includegraphics[width=0.3\linewidth]{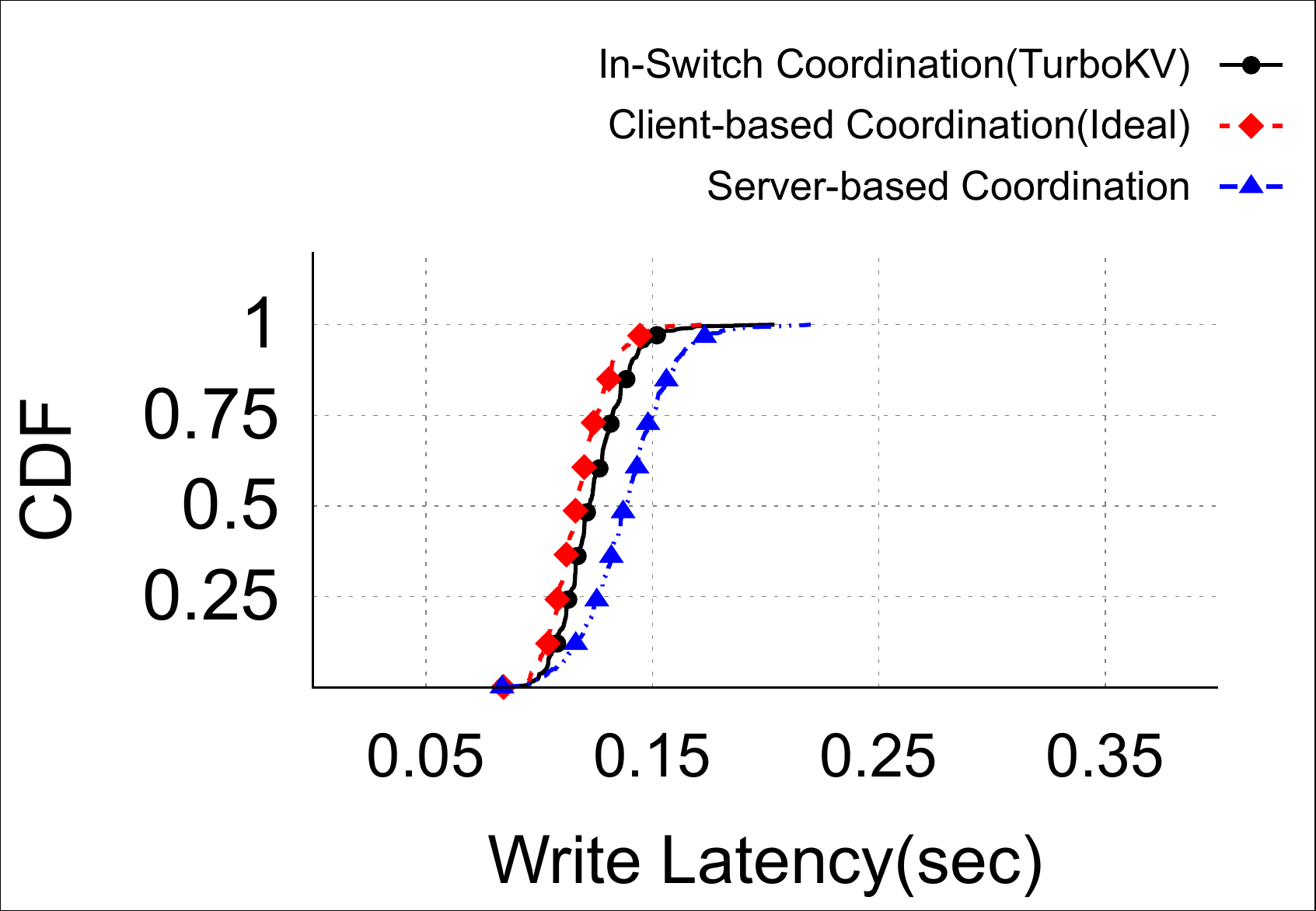}}
\hspace{0.03\linewidth}
\subfigure[Scan Latency]{\label{fig:scan_uniform}\includegraphics[width=0.3\linewidth]{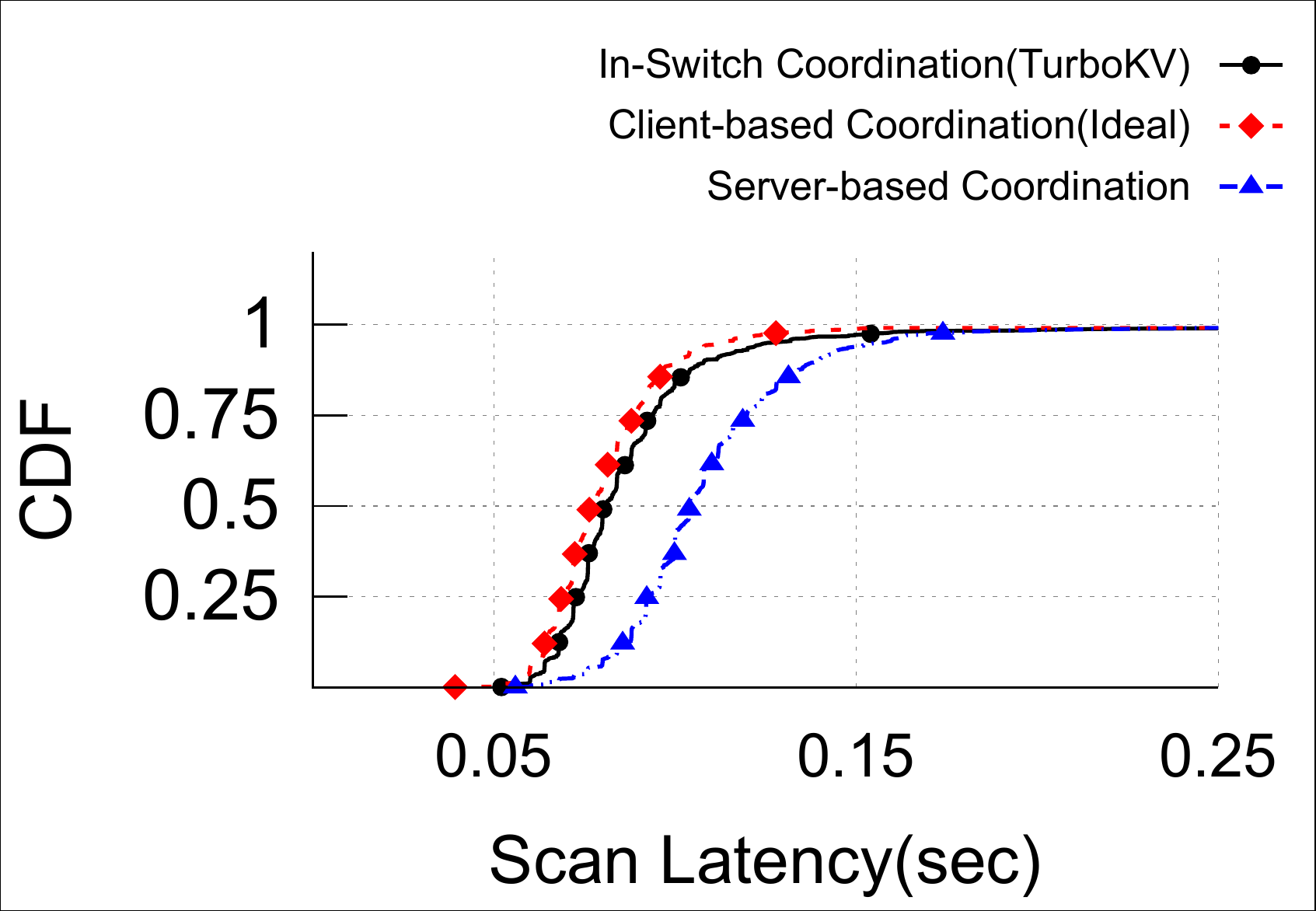}}
\caption{Key-value operations Latency for uniform Workload}
\label{fig:request_latency_uniform}
\end{figure*}

\begin{figure*}
\centering 
\subfigure[Read Latency]{\label{fig:read_zipf}\includegraphics[width=0.3\linewidth]{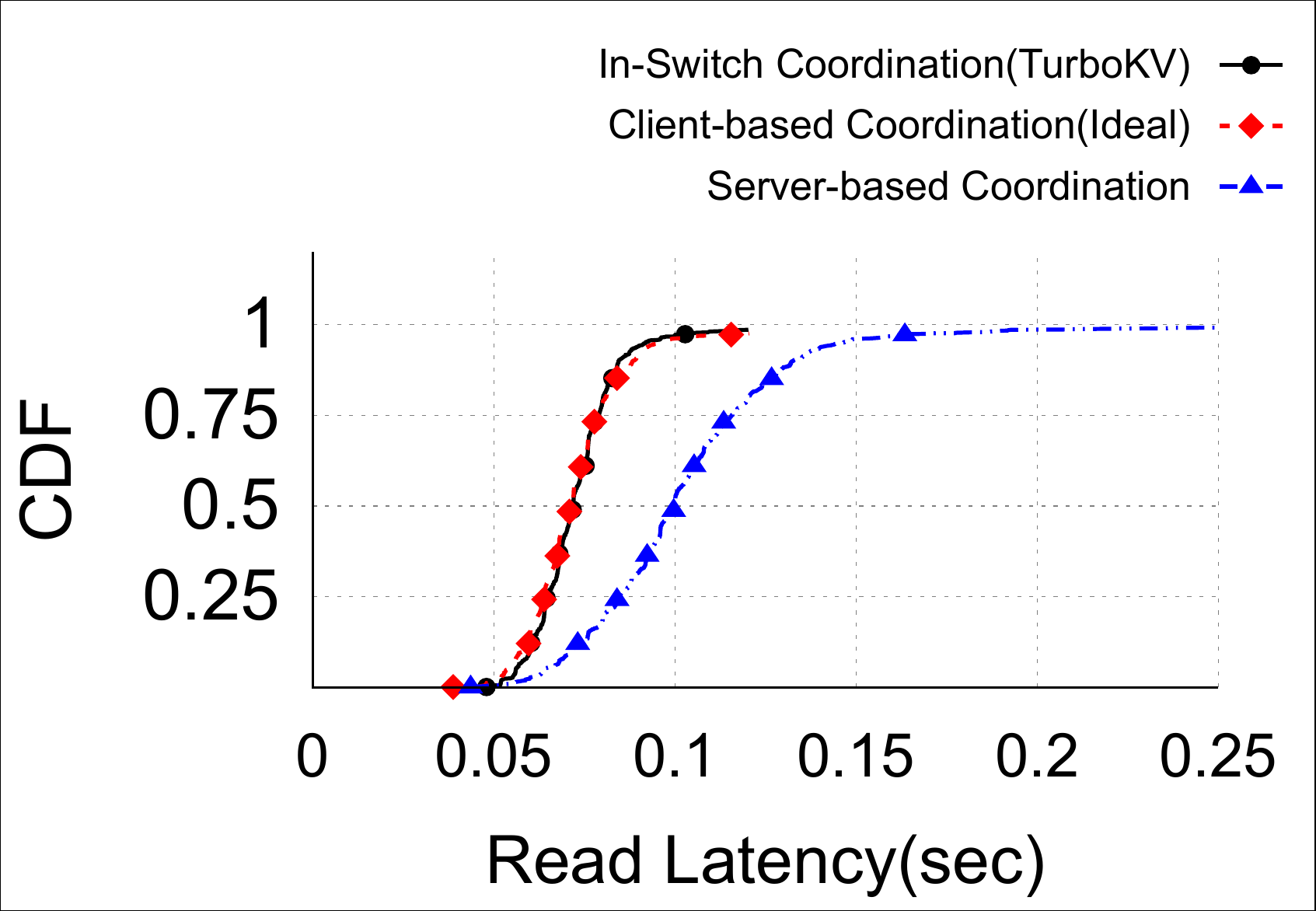}}
\hspace{0.03\linewidth}
\subfigure[Write Latency]{\label{fig:write_zipf}\includegraphics[width=0.3\linewidth]{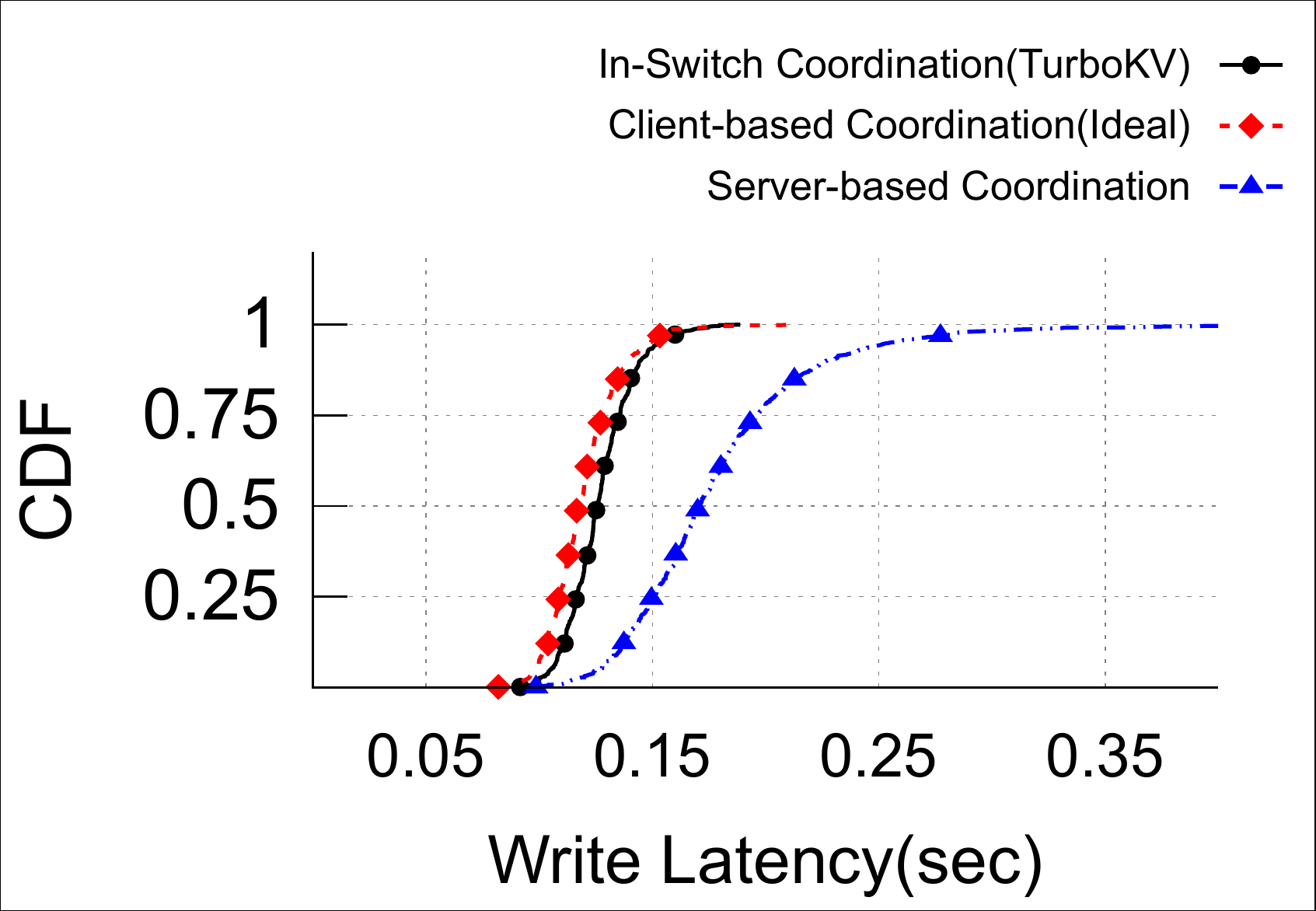}}
\hspace{0.03\linewidth}
\subfigure[Scan Latency]{\label{fig:scan_zipf}\includegraphics[width=0.3\linewidth]{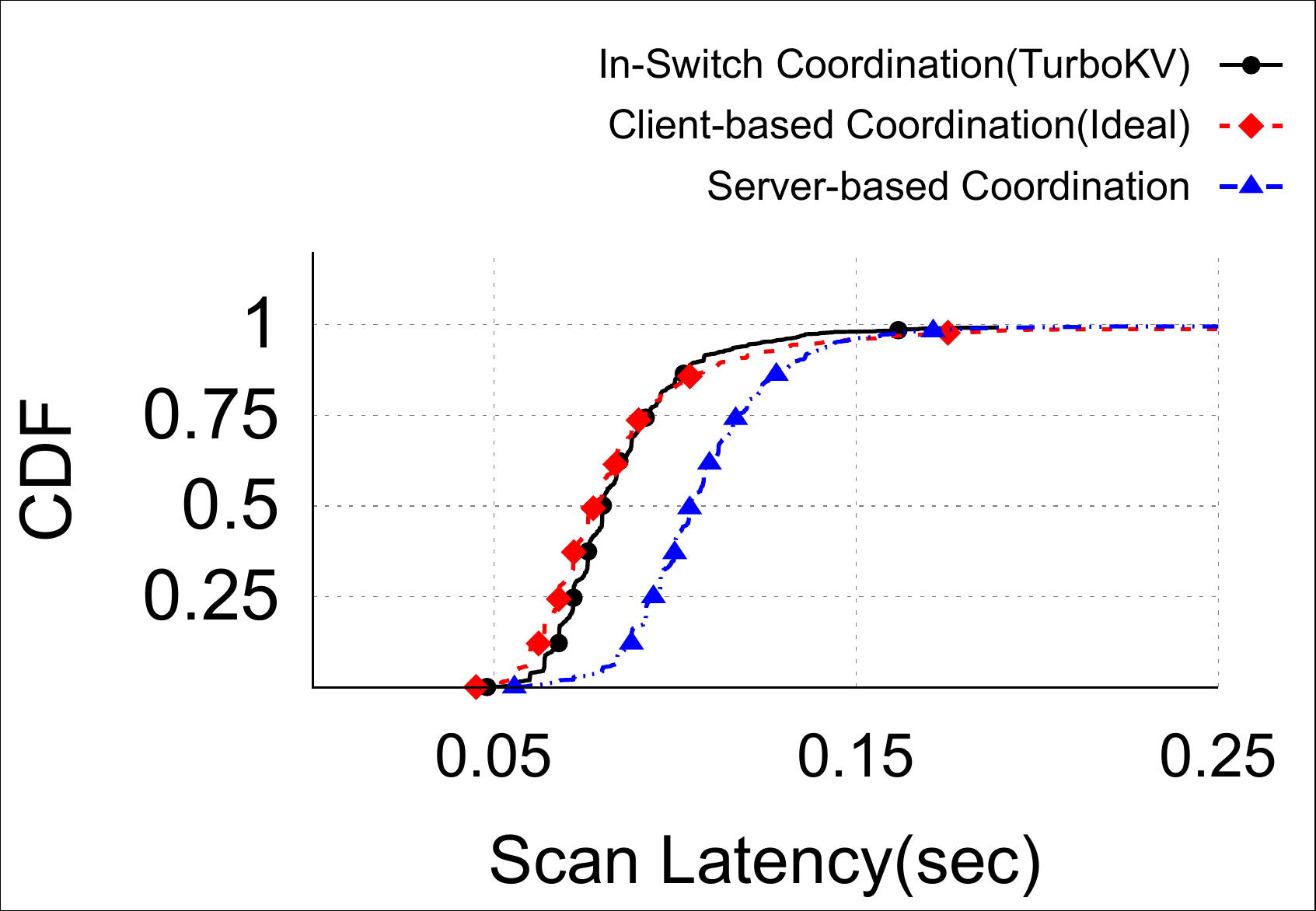}}
\caption{Key-value operations Latency for zipf-1.2 Workload}
\label{fig:request_latency_zipf}
\end{figure*}

\begin{table*}
\centering
\resizebox{\textwidth}{!}{\begin{tabular}{l|c|c|c|c|c|c|c|c|c|}
\cline{2-10}
                                                       & \multicolumn{3}{c|}{Read (Get) (msec)}                        & \multicolumn{3}{c|}{Write (Put) (msec)}                                                                 & \multicolumn{3}{c|}{Scan (Range) (msec)}                                                                \\ \cline{2-10} 
                                                       & Mean & 50th Percentile & \multicolumn{1}{l|}{99th Percentile} & \multicolumn{1}{l|}{Mean} & \multicolumn{1}{l|}{50th Percentile} & \multicolumn{1}{l|}{99th Percentile} & \multicolumn{1}{l|}{Mean} & \multicolumn{1}{l|}{50th Percentile} & \multicolumn{1}{l|}{99th Percentile} \\ \hline
\multicolumn{1}{|l|}{In-Switch Coordination (TurboKV)} & 72.5 & 71.2            & 103.3                                & 123.5                     & 121.8                                & 165.8                                & 84.3                      & 80                                 & 160.2                                 \\ \hline
\multicolumn{1}{|l|}{Client-driven Coordination}       & 69.8 & 68.8            & 98.6                                 & 117.5                     & 116.1                                & 153.5                                & 80.8                      & 76.5                                 & 139                                   \\ \hline
\multicolumn{1}{|l|}{Server-driven Coordination}       & 86.6 & 87.7            & 127.8                                & 138.2                     & 138                                  & 189                                  & 109                        & 104                                 & 184.5                                \\ \hline
\end{tabular}}
\caption{Request Latency Analysis under Uniform Workload}
\label{tab:uniform_latency}
\end{table*}

\begin{table*}
\centering
\resizebox{\textwidth}{!}{\begin{tabular}{l|c|c|c|c|c|c|c|c|c|}
\cline{2-10}
                                                       & \multicolumn{3}{c|}{Read (Get) (msec)}                         & \multicolumn{3}{c|}{Write (Put) (msec)}                                                                 & \multicolumn{3}{c|}{Scan (Range) (msec)}                                                                \\ \cline{2-10} 
                                                       & Mean  & 50th Percentile & \multicolumn{1}{l|}{99th Percentile} & \multicolumn{1}{l|}{Mean} & \multicolumn{1}{l|}{50th Percentile} & \multicolumn{1}{l|}{99th Percentile} & \multicolumn{1}{l|}{Mean} & \multicolumn{1}{l|}{50th Percentile} & \multicolumn{1}{l|}{99th Percentile} \\ \hline
\multicolumn{1}{|l|}{In-Switch Coordination (TurboKV)} & 72.2  & 71.8            & 105.3                                & 126.8                     & 125.4                                & 172.4                                & 87.3                      & 80.3                                   & 204                                \\ \hline
\multicolumn{1}{|l|}{Client-driven Coordination}       & 71.4  & 70.9            & 104                                  & 119.7                     & 117.2                                & 167.3                                & 85.6                      & 80                                 & 196                                 \\ \hline
\multicolumn{1}{|l|}{Server-driven Coordination}       & 102.8 & 99.8            & 206.8                                & 178.3                     & 170.9                                & 330.6                                & 112                      & 104.5                                 & 242.6                                \\ \hline
\end{tabular}}
\caption{Request Latency Analysis under Zipf-1.2 Workload}
\label{tab:zipf_latency}
\end{table*}

Figures~\ref{fig:request_latency_uniform} and \ref{fig:request_latency_zipf} show the CDF of all key-value operations latencies under uniform and Zipf-1.2 workloads, respectively, for \kvnet, ideal client-driven coordination and server-driven coordination. The analysis of these two figures is shown in \kvtable\ref{tab:uniform_latency} and \kvtable\ref{tab:zipf_latency}. Figures~\ref{fig:read_uniform}, \ref{fig:read_zipf}, \ref{fig:write_uniform}, and \ref{fig:write_zipf} show that the read and write latencies in \kvnet~are very close to the ideal client-driven coordination for uniform and skewed workload, because \kvnet~skips potential forwarding step like the client-driven coordination by managing the directory information in the switch itself. Compared to server-driven coordination, \kvnet~reduces the read latency by 16.3\% on the average and 19.2\% for the $99^{th}$ percentile for the uniform workload, and by 30\% on the average and 49\% for the $99^{th}$ percentile for the skewed workload. \kvnet~also reduces the write latency by 11\% on the average and 12.3\% for the $99^{th}$ percentile for the uniform workload, and by 29\% on the average and 48\% for the $99^{th}$ percentile for the skewed workload. The reduction in case of skewed workload is better than the reduction of uniform workload, because \kvnet~does not only skip the excess forwarding step but also removes the load from the storage nodes from being the request coordinator, and makes them focus only on answering the queries themselves, which reduces tail latencies at the storage nodes.

Figures~\ref{fig:scan_uniform}, and \ref{fig:scan_zipf} shows that \kvnet~reduces the scan latency by 23\% on average and 13\% for $99^th$ percentiles for uniform workloads, and by 22\% on average and 16\% for $99^th$ percentile for skewed workloads when compared with the server-driven coordination. But, when compared with the ideal client-driven coordination, \kvnet~increases the latency of the scan operation by range of 2 - 15\% for uniform and skewed workloads, because of the latency introduced inside the switch from packet circulation and cloning to divide the requested range when it spans multiple storage nodes.

\section{Related Work} \label{relatedwork}

\noindent\textbf{Distributed Key-value Stores.} Key-value storage is widely used to support lots of large-scale applications. Some key-value stores, e.g., Redis~\cite{redisref}, RAMCloud~\cite{ramcloud}, and memcached~\cite{memcached}, manage data in DRAM for faster data access. Other key-value stores, e.g., Dynamo~\cite{dynamo}, Cassendra~\cite{cassendra}, LevelDB~\cite{leveldb} and RocksDB~\cite{rocksdb} are presistent key-value stores which save data on presistent storage devices, while other key-value stores, e.g., AeroSpike~\cite{aerospike} use hybrid storage(DRAM and SSD). 
Distributing data over several key-value store instances has been widely studied. Some systems use hash functions to distribute the data among storage nodes. Dynamo~\cite{dynamo} and Cassandra~\cite{cassendra} use consistent hashing, while Redis~\cite{redisref} and Aerospike~\cite{aerospike} use a hash function to distribute data into several hashing slots. Other key-value stores, e.g., LevelDB~\cite{leveldb} and RocksDB~\cite{rocksdb}, use alternate approach to distribute the data, they use the key range partitioning where keys are on Sorted String Tables. \kvnet~supports hash partitioning of the data and range partitioning. Applications can decide the way to partition the data and the corresponding directory information will be stored in switches' data plane.

To achieve durability and high availability, data partitions are replicated over several storage nodes. Dynamo, Cassendra, Aerospike, and Redis replicate the data partitions over several storage nodes and save this information on a mapping table which is replicated also on all storage nodes. \kvnet~also replicates the data partitions over several storage nodes and follows the chain replication model to guarantee strong consistency. It stores the chain of each partition on the switch data plane. All existing key-value stores use either client-based coordination ~\cite{aerospike, dynamo}, server-based coordination~\cite{cassendra, redisref, dynamo} or a single elected node coordination~\cite{ramcloud} to deliver requests from clients to storage nodes. \kvnet~uses in-switch coordination with the key-based routing protocol to route requests from clients to storage nodes directly.  

\noindent\textbf{Hardware Acceleration.} Lots of work used hardware to speed up the performance of the distributed systems. NetPaxos~\cite{netpaxos1,netpaxos2} implements Paxos on switches. NetCache~\cite{netcache} implements a rack-scale on-switch cache, while DistCache~\cite{distcache} scales up the NetCache design to multiple racks inside the data center network. NetChain~\cite{netchain} uses the network switches to implement in-network key-value store, but it is bounded by the limited storage in the network switches. There is also SwitchKV~\cite{switchkv} which is a cluster scale system to balance the key value stores via caching. SwitchKV uses the OpenFlow switches to save a forwarding rule for each cached key-value pair to route the request to the right caching node. \kvnet~uses the switches to act as the request coordination nodes that save the partition management information along with the key-based-routing protocol to route the request to the target storage nodes. In \kvnet, the actual key-value pairs are saved on storage servers which makes it not limited to applications of small data sizes.
Other examples that use the hardware to increase the perfromance include, but not limited to, JoiNS~\cite{joins} which uses the OpenFlow switches to prioritize I/O packets to meet their latency SLO, KVDirect~\cite{kvdirect}, a high performance KVS that leverages programmable NIC to extend RDMA primitives and enable remote direct key-value access to the main host memory, iSwitch~\cite{iswitch} which uses the network switches to improve the performance of the distributed reinforcement learning, and Ibex~\cite{ibex} which supports advanced SQL offloading using FPGA.     
 
\section{Conclusion}\label{sec:conclusion}
In this paper, we presented \kvnet; a novel distributed key-value store architecture that leverages the power and flexibility of the new programmable switches. \kvnet~uses the in-switch coordination approach that utilizes the switches as partitions management nodes to store the key-value store partitions locations and replicas information along the path from clients to storage nodes.  The programmable switches use key-based routing to route packets from clients to storage nodes. \kvnet~decreases the query response time and improve system throughput. We believe that \kvnet~can be deployed on the programmable switches currently integrated in the data center's network to improve the performance of distributed key-value stores.

\bibliographystyle{plain}
\vspace{-6 px}
\bibliography{Distributed_KV_Store_with_Programmable_switches}

\end{document}